\documentclass[a4paper,11pt]{article}
\usepackage[T1]{fontenc}
\usepackage[utf8]{inputenc}

\usepackage{amsmath,amsthm,amssymb}
\usepackage{graphicx,subcaption,tikz}
\usepackage{comment,xspace,paralist}							                              
\usepackage[shortlabels]{enumitem}	
\usepackage{hyperref}       
\hypersetup{
    colorlinks=true,
    citecolor = blue,
    linkcolor=black,
    filecolor=magenta,
    urlcolor=cyan,
    bookmarks=true,
}
\usepackage{tabularx}
\usetikzlibrary{shapes,snakes}
\usepackage{thm-restate}
\usepackage[margin=1in]{geometry}
\usepackage{todonotes}
\usepackage{lmodern}

\newtheorem{theorem}{Theorem}
\newtheorem{lemma}[theorem]{Lemma}
\newtheorem{corollary}[theorem]{Corollary}
\newtheorem{claim}{Claim}[theorem]

\newtheorem{observation}[theorem]{Observation}

\theoremstyle{definition}

\newcommand{\is}{{\sc Independent Set}\xspace}
\newcommand{\sat}[1]{{\sc $#1$-Sat}\xspace}
\newcommand{\msat}[1]{{\sc Monotone $#1$-Sat}\xspace}
\newcommand{\col}[1]{{\sc $#1$-Coloring}\xspace}
\newcommand{\ext}[1]{{\sc $#1$-Precoloring Extension}\xspace}
\newcommand{\lcol}[1]{{\sc List $#1$-Coloring}\xspace}

\DeclareMathOperator{\girth}{girth}
\newenvironment{inproof}{\noindent {\emph{Proof of Claim.}}}{\hfill$\blacksquare$ \medskip}

\title{Complexity of $C_k$-coloring in hereditary classes of graphs\footnote{The extended abstract of the paper was presented on ESA 2019~\cite{DBLP:conf/esa/ChudnovskyHRSZ19}.}}

\author{Maria Chudnovsky\footnote{Princeton University, Princeton, NJ 08544, USA. Supported by NSF grant DMS-1763817. This material is based upon work supported in part by the U. S. Army Research Laboratory and the  U. S. Army Research Office under grant number W911NF-16-1-0404.}
\and Shenwei Huang\footnote{College of Computer Science, Nankai University, Tianjin 300350, China. Supported by the National Natural Science Foundation of China (12171256). The corresponding author (Email: shenweihuang@nankai.edu.cn).}
\and Pawe{\l} Rz\k{a}\.{z}ewski\footnote{Faculty of Mathematics and Information Science, Warsaw University of Technology, Warsaw, Poland. Supported by Polish National Science Centre grant no. 2018/31/D/ST6/00062.}
\and Sophie Spirkl\footnote{Rutgers University, Piscataway, NJ 08854, USA. This material is based upon work supported by the National Science Foundation under Award No. DMS-1802201. Current address: Department of Combinatorics and Optimization, University of Waterloo, Waterloo, Ontario N2L3G1.}
\and Mingxian Zhong\footnote{Lehman College and the Graduate Center, CUNY, Bronx, NY 10468, USA}}

\begin{document}

\maketitle

\begin{abstract} 
For a graph $F$, a graph $G$ is \emph{$F$-free} if it does not contain an induced subgraph isomorphic to $F$. 
For two graphs $G$ and $H$, an \emph{$H$-coloring} of $G$ is a mapping $f:V(G)\rightarrow V(H)$
such that for every edge $uv\in E(G)$ it holds that  $f(u)f(v)\in E(H)$. We are interested in the complexity of the problem $H$-{\sc Coloring}, which asks for the existence of an $H$-coloring of an input graph $G$.
In particular, we consider $H$-{\sc Coloring} of $F$-free graphs, where $F$ is a fixed graph and $H$ is an odd cycle of length at least 5. This problem is closely related to the well known open problem of determining the complexity of 3-{\sc Coloring} of $P_t$-free graphs.

We show that for every odd $k \geq 5$, the $C_k$-{\sc Coloring} problem, even in the list variant, can be solved in polynomial time in $P_9$-free graphs. The algorithm extends to the list version of $C_k$-{\sc Coloring}, where $k \geq 10$ is an even number.

On the other hand, we prove that if some component of $F$ is not a subgraph of a subdivided claw, then the following problems are NP-complete in $F$-free graphs:
\begin{enumerate}[a)]
\item the precoloring extension version of $C_k$-{\sc Coloring} for every odd $k \geq 5$;
\item the list version of $C_k$-{\sc Coloring} for every even $k \geq 6$.
\end{enumerate}

\end{abstract}
\newpage

\section{Introduction}
For graphs $G$ and $H$, a {\em homomorphism from} $G$ {\em to} $H$ is a mapping $f:V(G)\rightarrow V(H)$
such that $f(u)f(v)\in E(H)$ for every edge $uv\in E(G)$. 
An {\em isomorphism from} $G$ {\em to} $H$ is a bijection $f:V(G)\rightarrow V(H)$ such that $f(u)f(v)\in E(H)$ if and only if $uv\in E(G)$. 
It is straightforward to see that if $H$ is a complete graph with $k$ vertices, then every homomorphism to $H$ is in fact a $k$-coloring of $G$ (and vice versa). This shows that homomorphisms can be seen as a generalization of graph colorings. Because of that, a homomorphism to $H$ is often called an {\em $H$-coloring}, and vertices of $H$ are called {\em colors}. We also say that $G$ is {\em $H$-colorable} if $G$ has an $H$-coloring. 

In what follows, the target graph $H$ is always fixed. We are interested in the complexity of the following computational problem, called \col{H}.

\medskip
\noindent\begin{tabularx}{\textwidth}{| X |}
\hline
{\bf Problem:  \col{H}} \\
{\bf Instance:} A graph $G$.\\
{\bf Question:} Does there exist a homomorphism from $G$ to $H$?\\
\hline
\end{tabularx}

\subsubsection*{Complexity of variants of \col{H}}

Since \col{H} is a generalization of \col{k}, it is natural to try to extend results for \col{k} to target graphs $H$ which are not complete graphs. For example, it is well-known that \col{k} enjoys a complexity dichotomy: it is polynomial-time solvable if $k\le 2$, and NP-complete otherwise. 
The complexity dichotomy for \col{H} was described by Hell and Ne\v{s}et\v{r}il in their seminal paper~\cite{HN90}: they  proved that the problem is polynomial-time solvable if $H$ is bipartite, and NP-complete otherwise.

Since then, there have been numerous studies on variants of \col{H}.  Let us mention two of them.
In the \ext{H} problem, we are given a graph $G$, a subset $W\subseteq V(G)$ and a mapping $h:W\rightarrow V(H)$. The problem is to decide if $h$ can be extended to an $H$-coloring of $G$, that is, if there is an $H$-coloring $f$ of $G$ such that $f|W=h$. In the \lcol{H} problem, the input consists of a graph $G$ with  an \emph{$H$-list assignment}, which is a function $L: V(G) \rightarrow 2^{V(H)}$ that assigns a subset of $V(H)$ to each vertex of $G$. We ask  if there is an \emph{$L$-coloring}, that is, an $H$-coloring $f$ of $G$ such that $f(v)\in L(v)$ for each $v\in V(G)$. In such a case we say that $(G,L)$ is {\em $H$-colorable}. We formulate these two problems as follows.

\medskip
\noindent\begin{tabularx}{\textwidth}{| X |}
	\hline
	{\bf Problem: \ext{H}}\\
	{\bf Instance:} A graph $G$, a subset $W\subseteq V(G)$, and a mapping $h \colon W \to V(H)$.\\
	{\bf Question:} Can $h$ be extended to an $H$-coloring of $G$?\\
	\hline
\end{tabularx}

\medskip

\noindent\begin{tabularx}{\textwidth}{| X |}
	\hline
	{\bf Problem: \lcol{H}}\\
	{\bf Instance:} A pair $(G,L)$ where $G$ is a graph and $L$ is an $H$-list assignment.\\
	{\bf Question:} Does there exist an $H$-coloring of $(G,L)$?\\
	\hline
\end{tabularx}
\medskip

When $H=K_k$, we write \ext{k} and \lcol{k} for \ext{K_k} and \lcol{K_k}, respectively.
Clearly \ext{H} can be seen as a restriction of \lcol{H} in which every list is either a singleton, or contains all vertices of $H$. This is the reason why it is sometimes called {\em one-or-all list homomorphism (coloring) problem}~\cite{DBLP:journals/combinatorica/FederHH99}.

Note that \lcol{H} is at least as hard as \ext{H}, which in turn is at least as hard as \col{H}.
Thus, any algorithmic result for \lcol{H} carries over to the other two problems, and any hardness result
for \col{H} carries over to \ext{H} and \lcol{H}.

The complexity dichotomy for \lcol{H} was proven in three steps: first, for reflexive graphs $H$~\cite{FEDER1998236}, then for irreflexive graphs $H$~\cite{DBLP:journals/combinatorica/FederHH99}, and finally for all graphs $H$~\cite{DBLP:journals/jgt/FederHH03}.
In general, variants of \col{H} can be seen in a wider context of Constraint Satisfaction Problems (CSP). A full complexity dichotomy for this family of problems has been a long-standing open question, known as the {\em CSP dichotomy conjecture} of Feder and Vardi~\cite{DBLP:conf/stoc/FederV93}. After a long series of partial results, the problem was finally solved very recently, independently by Bulatov~\cite{DBLP:conf/focs/Bulatov17} and by Zhuk~\cite{DBLP:conf/focs/Zhuk17}.

A natural approach in dealing with computationally hard problems is to consider restricted instances, in the hope of understanding the boundary between easy and hard cases. For example, 
it is known that \col{H} can be solved in polynomial time for perfect graphs, because it suffices to test whether $\omega(G)>\omega(H)$, which can be done
in $O(|V(G)|^{|V(H)|})$ time. If $\omega(G)>\omega(H)$, then the answer is no, as there is no way to map the largest clique of $G$ to $H$. Otherwise the answer is yes, since $\omega(G)$-coloring of $G$ can be translated to a homomorphism of $G$ to the largest clique of $H$, and thus to $H$.
The situation changes when we consider the more general setting of \ext{H} and \lcol{H}.
For any fixed graph $H$, \lcol{H} (and thus \ext{H} and \col{H}) can be solved in polynomial time for input graphs with bounded tree-width. 
Combining this with an observation that any graph with a clique larger than $\omega(H)$ has no $H$-coloring, 
we obtain polynomial-time algorithms for chordal graphs~\cite{FHKN05}. 
For permutations graphs, \lcol{H} can also be solved in polynomial time via a recursive branching algorithm~\cite{EST14}. 
For bipartite input graphs, however, \ext{3} (i.e., \ext{K_3}) is already NP-complete~\cite{Kr93}.
Other restricted inputs have been studied too, e.g.\ bounded-degree graphs~\cite{DBLP:journals/dm/GalluccioHN00,DBLP:journals/dam/FederHH09}. 
For more results on graph homomorphisms, we refer to the monograph by Hell and Ne\v{s}et\v{r}il~\cite{HN04}.

In this paper, we study the complexity of \col{H} for minimal non-bipartite graphs $H$, namely when $H$ is an odd cycle. We also extend these results to the \lcol{H} problem, also for even target cycles.

Although \col{H} is well-studied for hereditary classes characterized by infinitely many forbidden induced subgraphs~\cite{EST14,FHKN05,Kr93}, 
not much is known for hereditary classes characterized by a finite set of forbidden induced subgraphs~\cite{DBLP:journals/dam/FederHH09,DBLP:journals/dm/GalluccioHN00}. 
Here we  initiate such a study for hereditary classes characterized by a single forbidden induced subgraph.

\subsubsection*{Graphs with forbidden induced subgraphs}

A rich family of restricted graph classes comes from forbidding some small substructures. For graphs $G$ and $F$, 
we say that $G$ {\em contains} $F$ if $F$ is an induced subgraph of $G$. By {\em $F$-free graphs} 
we mean the class of graphs that do not contain $F$. Note that this class is {\em hereditary}, that is, it is closed under taking induced subgraphs.

The complexity of  \col{k} for hereditary graph classes has received much attention in the past two decades
and significant progress has been made. Of particular interest is the class of $F$-free graphs for a fixed graph $F$.
For any fixed $k\ge 3$, the \col{k} problem remains NP-complete for $F$-free graphs whenever $F$ is not a linear forest (a collection of disjoint paths)~\cite{Ho81,KL07}. The simplest linear forests are paths, and the complexity of \col{k} in $P_t$-free graphs has been studied by many researchers. 

On the positive side, Ho\`{a}ng, Kami\'{n}ski, Lozin, Sawada, and Shu~\cite{HKLSS10}
gave a recursive algorithm showing that \col{k} can be solved in polynomial time for $P_5$-free graphs for any fixed $k$.
Bonomo, Chudnovsky, Maceli, Schaudt, Stein, and Zhong~\cite{BCMSSZ17} showed that \col{3} can be solved in polynomial time in $P_7$-free graphs. Moreover, very recently, Chudnovsky, Spirkl, and Zhong proved that \col{4} is polynomial-time solvable in $P_6$-free graphs~\cite{DBLP:journals/corr/abs-1802-02282,DBLP:journals/corr/abs-1802-02283,DBLP:conf/soda/SpirklCZ19}. 

On the negative side, Woeginger and Sgall~\cite{WS01} demonstrated the NP-completeness
of \col{5} for $P_8$-free graphs and \col{4} for $P_{12}$-free graphs. Later on, these NP-completeness results
were improved by various researchers and the strongest result is due to Huang~\cite{Hu16} who proved that
\col{4} is NP-complete for $P_7$-free graphs and \col{5} is NP-complete for $P_6$-free graphs.
These results settle the complexity of \col{k} for $P_t$-free graphs for all pairs $(k,t)$, except for the complexity of \col{3} for $P_t$-free graphs when $t\ge 8$. Interestingly, all polynomial-time results carry over to the list variant, except for the case of \lcol{4} of $P_6$-free graphs, which was shown to be NP-complete by Golovach, Paulusma, and Song~\cite{DBLP:journals/iandc/GolovachPS14}. We refer the reader to the survey by Golovach, Johnson, Paulusma, and Song~\cite{DBLP:journals/jgt/GolovachJPS17} for more information about coloring graphs with forbidden subgraphs.

Understanding the complexity of \col{3} in $P_t$-free graphs seems a hard problem -- on the one hand, algorithms even for small values of $t$ are difficult to construct, and on the other hand all our hardness reductions appear to introduce long induced paths. Let us mention a problem whose complexity is equally elusive: \is. Alekseev~\cite{alekseev1982effect} observed that \is is NP-complete in $F$-free graphs whenever $F$ is not a path or a subdivided claw. For $P_t$-free graphs, polynomial-time algorithms are known only for small values of $t$: currently, the best result is the recent polynomial-time algorithm for $P_6$-free graphs by Grzesik, Klimo\v{s}ova, Pilipczuk, and Pilipczuk ~\cite{DBLP:journals/corr/GrzesikKPP17,DBLP:conf/soda/GrzesikKPP19}. On the other hand, the problem is not known to be NP-hard for any fixed $t$.

A natural question to ask is if the similar behavior of \col{3} and \is in $P_t$-free graphs is a part of a more general phenomenon. Recently, Groenland, Okrasa, Rz\k{a}\.{z}ewski, Scott, Seymour, and Spirkl~\cite{GORSSS18} shed some light on this question by showing that if $H$ does not contain two vertices with two common neighbors, then a very general, weighted variant of \col{H} can be solved  in time  $2^{O(\sqrt{tn\log n})}$ for $P_t$-free graphs. Clearly $K_3$ does not have two vertices with two common neighbors. Moreover, \is can be expressed as a weighted homomorphism to \begin{tikzpicture}[scale=.7, every node/.style={draw,circle,fill=white,inner sep=1pt,minimum size=5pt}]
\tikzset{every loop/.style={}}
\draw[line width=1] (0,0) -- (1,0);
\node at (0,0) {} edge [line width=1,in=140,out=40,loop] ();
\node at (1,0) {};
\end{tikzpicture}, which has the same property, and thus, for every $t$, both \col{3} and \is can be solved in subexponential time in $P_t$-free graphs (we note that a subexponential algorithm for \is in $P_t$-free graphs was known before~\cite{DBLP:journals/algorithmica/BacsoLMPTL19}). This implies that if one attempts to prove NP-completeness of any of these problems in $P_t$-free graphs, then, assuming the Exponential Time Hypothesis \cite{ImpagliazzoPaturi,DBLP:journals/jcss/ImpagliazzoPZ01}, such a reduction should be sufficiently complicated to introduce at least a quadratic blow-up of the instance.

In this paper, we study the complexity of variants of \col{H} when $H$ is a cycle.
Recall that \col{K_3} is equivalent to \col{3}, which is a well-studied problem.
Furthermore, even \lcol{C_4} is polynomial-time solvable in general graphs~\cite{DBLP:journals/combinatorica/FederHH99}.
Thus we focus on the case that $H=C_k$ for $k \geq 5$.
Note that by the result of Groenland {\em et al.} \cite{GORSSS18}, this problem can be solved in subexponential time in $P_t$-free graphs. We are interested in better classification of polynomial and NP-hard cases.

\subsubsection*{Our contribution} 

First, we show that \lcol{C_k} can be solved in polynomial time in $P_9$-free graphs for every $k\geq 9$ or $k=5$ or $7$.

\begin{restatable}{theorem}{thmpolynomial}\label{thm:polynomial}
Let $k\geq 9$ or $k=5$ or $7$, and let $(G,L)$ be an instance of \lcol{C_k} where $G$ is a $P_9$-free graph of order $n$. 
	Then one can determine in $O(n^{12k+3})$ time if $(G,L)$ is $C_k$-colorable, 
	and find a $C_k$-coloring of $(G,L)$ if one exists.  
\end{restatable}
The algorithm is described in detail in Section \ref{sec:poly}. It builds on the recent work on \col{3} $P_7$-free graphs~\cite{BCMSSZ17}. The high-level idea of the algorithm is the following: First, we partition the graph into a so-called {\em layer structure} and guess the colors of a constant number of vertices. This precoloring propagates to other vertices, reducing their lists. We keep guessing the colors of other vertices, transforming the input instance $(G,L)$ into a set of $n^{O(k)}$ subinstances, such that both the following conditions are satisfied:
\begin{enumerate}[(i)]
	\item $(G,L)$ admits a $C_k$-coloring if and only if one of these subinstances admits a $C_k$-coloring;
	\item each subinstance can be solved in polynomial time by a reduction to \sat{2}.
\end{enumerate}

In Section \ref{sec:polyimp}, we show that the above algorithm can be improved such that it remains polynomial when $k$ is part of the input. In particular, we prove the following theorem.
\begin{restatable}{theorem}{thmimprove}\label{thm:improve}
	Let $k\geq 9$ or $k=5$ or $7$, and let $(G,L)$ be an instance of \lcol{C_k} where $G$ is a $P_9$-free graph of order $n$. Then one can determine in $O(k \cdot n^{111})$ time if $(G, L)$ is colorable, and find such a $C_k$-coloring  if one exists. 
\end{restatable}
In Section \ref{sec:NPc}, we study the complexity of variants of \col{H} in $F$-free graphs and prove the following theorem.
\begin{restatable}{theorem}{thmhard}\label{thm:main-hardness}
Let $F$ be a connected graph. If $F$ is not a subgraph of a subdivided claw, then for every odd $k \geq 5$ 
the \ext{C_k} problem is NP-complete for $F$-free graphs.
\end{restatable}

We prove the theorem in several steps, each dealing with certain class of ``hard'' graphs $F$.
In most cases, we actually prove hardness for the more restricted \col{C_k} problem.

Then we turn our attention to \lcol{C_k} in $F$-free graphs, when $k$ is even. It is known that for  general graphs this problem is polynomial time solvable for $k=4$, and NP-complete for every $k \geq 6$~\cite{DBLP:journals/combinatorica/FederHH99}. We prove the following analogue of \autoref{thm:main-hardness}. 

\begin{restatable}{theorem}{thmevenhard}\label{thm:main-hardness-even}
Let $F$ be a connected graph. If $F$ is not a subgraph of a subdivided claw, then for every even $k \geq 6$ 
the \lcol{C_k} problem is NP-complete for $F$-free graphs.
\end{restatable}

Observe that the statements of \autoref{thm:main-hardness} and \autoref{thm:main-hardness-even} are similar to the previously mentioned result of Alekseev for \is~\cite{alekseev1982effect}. 

Finally, in \autoref{sec:conclusion}, we state some open questions for future research.

\section{Preliminaries}\label{sec:pre}
Let $G$ be a simple graph. For $X \subseteq V(G)$, we denote by $G|X$ the subgraph induced by $X$, and denote by
$G \setminus X$ the graph $G|(V(G) \setminus X)$.
When $G$ is clear from the context and it does not lead to confusion, we use induced subgraphs and their  vertex sets interchangeably.
In particular, we say that $X$ is {\em connected} if $G|X$ is connected.
A set $X \subseteq V(G)$ is \emph{stable} or \emph{independent} if $G|X$ is an edgeless graph.
For two disjoint subsets $A,B \subset V(G)$, we say that $A$ is \emph{complete} to $B$ if every vertex of $A$ is adjacent to every vertex of $B$, and that $A$ is \emph{anticomplete} to $B$ if every vertex of $A$ is nonadjacent to every vertex of $B$. If $A=\{a\}$ we write $a$ is complete (or anticomplete) to $B$ to mean that $\{a\}$ is complete (or anticomplete) to $B$.
For $X \subseteq V(G)$, we say that $e \in E(G)$ is \emph{an edge of $X$} if both endpoints of $e$ are in $X$.
For $v \in V(G)$ we write $N_G(v)$ (or $N(v)$ when there is no danger of confusion) to mean the set of vertices of $G$ that are adjacent to  $v$. 
Observe that since $G$ is simple, $v \not \in N(v)$. For $X \subseteq V(G)$ we define
$N(X)=\left(\bigcup_{v \in X} N(v)\right) \setminus X$. We say that the set $S$ \emph{dominates} $X$, or $S$ is a \emph{dominating set} of $X$ if $X\subseteq S\cup N(S)$. We write that $S$ dominates $G$ when we mean that it dominates $V(G)$.
A component of $G$ is {\em trivial} if it has only one vertex and {\em nontrivial} otherwise.

We use $[k]$ to denote the set $\{1, 2, \dots, k\}$.
We denote by $P_t$ the path with $t$ vertices.
A \emph{path in a graph  $G$} is  a sequence $v_1-\cdots -v_t$ of pairwise 
distinct vertices such that for any $i,j\in [t]$, $v_iv_j\in E(G)$ if and only if $|i-j|=1$.  The {\em length} of this path is $t-1$.  
We denote by $V(P)$ the set $\{v_1, \ldots, v_t\}$. If $a, b \in V(P)$, say $a=v_i$ and $b=v_j$ with $i<j$, then $a-P-b$ is the 
path $v_i-v_{i+1}-\cdots -v_j$, and $b-P-a$ is the path $v_j-v_{j-1}-\cdots-v_i$.

 

Let $k \geq 3$ be an odd integer. We denote by $C_k$ a cycle with $k$ vertices $1,2,\ldots,k$ that appear along the cycle in this order.
The calculations on vertices of $C_k$ will be preformed modulo $k$, with 0 interpreted as the vertex $k$.

We say that $(G,L')$ is a {\em subinstance} of $(G,L)$ if $L'(v)\subseteq L(v)$ for every $v\in V(G)$.
Two $C_k$-list assignments $L$ and $L'$ of $G$  are {\em equivalent} if $(G,L)$ is $C_k$-colorable if and only if $(G,L')$ is $C_k$-colorable.
A $C_k$-list assignment $L$ is {\em equivalent} to a set $\mathcal{L}$ of $C_k$-list assignments of a graph $G$ if there is $L' \in \mathcal{L}$ such that $(G,L)$ is equivalent to $(G,L')$. 

Let $(G,L)$ be an instance of \lcol{C_k}. All indices below are computed modulo $k$. For an edge $vw\in E(G)$, we {\em update} $v$ {\em from} $w$ if one of the following is performed.
\begin{compactitem}
	\item If $L(w)=\{i\}$  for some $i\in[k]$, then replace the list of $v$ by $\{i-1,i+1\}\cap L(v)$.
	\item If $L(w)=\{i-1,i+1\}$ for some $i\in[k]$,  then replace the list of $v$ by $\{i,i+2,i-2\} \cap L(v)$.
	\item If $L(w)=\{i,i-2,i+2\}$ , $L(v)=\{j,j+2,j-2\}$ for some $i,j\in[k]$, 
	then replace the list of $v$ by $\{i-1,i+1,i-3,i+3\}\cap L(v)$.
\end{compactitem}


Clearly, any update creates an equivalent subinstance of $(G,L)$. Note that in the graph homomorphism literature this operation is usually referred to as {\em edge (or arc) consistency} and it is performed in the beginning of most algorithms solving variants of \col{H}~\cite{HN04,DBLP:journals/dm/LaroseL13}. However, we keep the name ``update'' to emphasize that we will only perform it at certain points in our algorithm.
We say that an update of $v$ from $w$ is {\em effective} if the size of the list of $v$ decreases by at least 1,
and {\em ineffective} otherwise.
Note that an update is effective if and only if 
there exists an element $c\in L(v)$ which is not an element of $\{i-1,i+1\}$, $\{i,i+2,i-2\}$ or $\{i-1,i+1,i-3,i+3\}$ 
depending on the case in the definition of an update.

Let $A$ be an induced subgraph of $G$. A $C_k$-list assignment $L$ is said to be {\em reduced} on $A$ if no effective update can be performed in $A$. 
It is well-known that one can obtain a reduced list assignment in polynomial time. We include a proof below for the sake of completeness.

\begin{lemma}\label{lem:reduced list}
	Let $G$ be a graph of order $n$, $L$ be a $C_k$-list assignment and $A$ be an induced subgraph of $G$. 
	There exists an $O(n^3)$-time algorithm to obtain an equivalent subinstance $(G,L')$ of $(G,L)$ such that $L'$ is reduced  on $A$
	or determine that $(G,L)$ has no $C_k$-coloring.
\end{lemma}

\begin{proof}
	We obtain $L'$ by performing updates exhaustively. 
	For each edge $vw\in E(A)$, we check if the update of $v$ from $w$ is effective. 
	If so, we perform the update. 
	If the list of $v$ becomes empty after the update, then we stop and claim that $(G,L)$ is not $C_k$-colorable.
	Otherwise we repeat until no effective update can be found.
	
	Since any single update results in an equivalent subinstance, we obtain a subinstance that is equivalent to $(G,L)$ at the end. Let $(G,L')$ be the subinstance we produce at the end, then it is clear that  $L'$ is reduced on $A$.
	It takes $O(n^2)$ time to find an effective update and $O(1)$ time to perform such a update.
	Moreover, since each effective update decreases the list of a vertex by at least 1 and $\sum_{v\in V(G)}|L(v)|\le kn$,
	effective updates can be performed at most $kn$ times. Thus, the total running time is $O(n^3)$.
\end{proof}

We now introduce two more tools that are important for our purpose. The first one is purely graph-theoretic and describes the structure of $P_t$-free graphs.

\begin{theorem}[\cite{CS16}]\label{thm:Pt}
Let $G$ be a connected $P_t$-free graph with $t\ge 4$. Then $G$ has a connected dominating set $D$ such that $G|D$ is either $P_{t-2}$-free
or isomorphic to $P_{t-2}$.
\end{theorem}

The next observation generalizes the observation by Edwards~\cite{Ed86} that \lcol{k} can be solved in polynomial time, whenever the size of each list is at most two. This was already noted by e.g. Feder and Hell~\cite{FEDER1998236} for the case when $A$ is $\emptyset$.

\begin{theorem}\label{thm:2SAT}
	Let $(G,L)$ be an instance of \lcol{C_k} where $G$ is of order $n$ and
	$A\subseteq V(G)$ is a stable set in $G$ satisfies the following:
	\begin{itemize}
		\item For every $v\in V(G)\setminus A$, $|L(v)|\le 2$.
		\item For every $v\in A$, $L(v)=\{i,i-2,i+2\}$ for some $i\in [k]$ and for every $u\in N(v)$, $L(u)$ is a subset of  $ \{i+1,i-1\},\{i+1,i-3\},\{i-1,i+3\}$ or $\{i+3,i-3\}$. 
	\end{itemize}  Then one can determine
	in $O(n^3)$ time if $(G,L)$ is $H$-colorable and find an $H$-coloring if one exists.
\end{theorem}

\begin{proof}
	If $L(v)=\emptyset$ for some $v\in V(G)$, then we claim that $(G,L)$ is not $C_k$-colorable.
	Otherwise we construct a \sat{2} instance as follows. 
	\begin{compactitem}
		\item For every $v\in V(G)\setminus A$ and $x \in L(v)$, we introduce a variable $v_x$.
		The meaning of $v_x$ is that $v_x$ is true if and only if $v$ is colored with color $x$.
		
		\item For every $v\in V(G)\setminus A$, we add a clause $\{v_x\}$ if $L(v)=\{x\}$, and we add two clauses
		$\{v_x,v_y\}$ and $\{\neg v_x, \neg v_y\}$ if $L(v)=\{x,y\}$. This ensures that the vertex $v$ gets exactly one color from $L(v)$.
		
		\item For every edge $uv\in E(G\setminus A)$ and every $x \in L(u)$, $y \in L(v)$, such that $y-x\not\equiv\pm 1 \mod k$, 
		we add a clause $\{\neg  u_x, \neg v_y\}$. This ensures that the edge $uv$ properly $C_k$-colored.
		
		\item For every vertex $v\in A$, for every $\{a,b\}\in N(v)$ and for every $x \in L(a)$, $y \in L(b)$ such that $x-y\not\equiv 0$ or $\pm 2 \mod k$, we add a clause $\{\neg  u_x, \neg v_y\}$.  Note that this way we forbid colorings of $a,b$ for which there is no choice of color for $v$. This is meant to ensure that the $C_k$-coloring of $V \setminus A$ can be extended to the vertices of $A$.
	\end{compactitem}
	
	Obviously if $(G,L)$ has an $C_k$-coloring, then the \sat{2} instance is satisfiable. In case the \sat{2} instance is satisfiable, we can obtain an $C_k$-coloring of $G\setminus A$ by setting $c(v)=x$ if $v_x$ is true for every $v\in V(G)\setminus A$ and every $x\in L(v)$. For each $v\in A$ with list $\{i,i-2,i+2\}$, $\bigcup_{u \in N(v)}L(v)\subseteq \{i+3,i+1,i-1,i-3\}$. Let $N=\bigcup_{u \in N(v)}\{c(u)\}$. By the clauses we posed, there are at most one of $\{i-1,i+3\}$, $\{i+1,i-3\}$ and $\{i-3,i+3\}$ in $N$ (recall that $k\neq 6,8$). It follows that $N$ is a subset of $\{i+1,i-1\}$, $\{i+1,i+3\}$ or $\{i-1,i-3\}$. We can assign $c(v)=i,i+2,i-2$ accordingly in each cases. It follows that $c$ extends to a $C_k$-coloring of $G$.
	The \sat{2} instance has $O(n)$ variables and $O(n^3)$ clauses and so it can be solved in $O(n^3)$ time by \cite{APT79}.
\end{proof}

\section{Polynomial algorithm for $P_9$-free graphs}\label{sec:poly}
Let $k=5,7$ or an integer greater than 8. In this section, we show that \lcol{C_k} can be solved in polynomial time for $P_9$-free graphs.


\noindent {\bf Outline of the proof.} 
The overall strategy is to reduce the instance $(G,L)$, in polynomial time, to polynomially many instances of \sat{2} in such a way that
$(G,L)$ is $C_k$-colorable if and only if at least one of the \sat{2} instances is a yes-instance. We then apply  \cite{APT79}
to solve each \sat{2} instance in polynomial time. 

More specifically, our algorithm, at a high level,  has the following five phases.
First, we apply \autoref{thm:Pt} to show that $G$ has a four-layer structure $\mathcal{P}=(S,X,Y,Z)$ such that the sets
$S$, $X$, $Y$ and $Z$ form a partition of $V(G)$, and  $S$ is connected and of bounded size. The set $S$ is called the {\em seed}.
Second, we branch on every possible coloring of $G|S$ that respects the lists $L$. For each of these colorings of $G|S$, 
we propagate the coloring on $S$ to the vertices of $G\setminus S$ via updates. 
After updating, the vertices in $S\cup X$ will have lists of size at most 2, but the vertices in $Y\cup Z$ may still have lists of size  more than 2. 
In the third step, we reduce the instance to polynomially many subinstances via branching in such a way that each of the subinstances avoids certain
configurations, which we call {\em bad paths}. 
Moreover, we ensure that the original instance is a yes-instance if and only if at least one of the new subinstances is a yes-instance.
Finally, using the fact that the subinstance has no bad paths, we may reduce the list size of vertices in $Y\cup Z$
and thus obtain an equivalent instance of \sat{2} using \autoref{thm:2SAT}. 

We now give a formal proof of the theorem.

\thmpolynomial*
\begin{proof}
\setcounter{theorem}{1}
We may assume that $G$ is connected, for otherwise we can solve the problem for each connected component of $G$ independently.
Moreover, one can determine in $O(n^3)$ time if $G$ has a triangle. If so, we stop and claim that $(G,L)$ is not $C_k$-colorable since
there is no homomorphism from a triangle to $C_k$ when $k\ge 4$. Therefore, we assume that $G$ is triangle-free from now on.

\vspace*{0.3cm}

\noindent {\bf Phase I.  Obtaining a layer structure.}

\begin{claim}\label{seed}
There exists $S\subseteq V(G)$ such that $|S|\leq 7$, $G|S$ is connected and $S\cup N(S)\cup N(N(S))$ dominates $G$. 
\end{claim}

\begin{inproof}
We apply \autoref{thm:Pt} to $G$: $G$ has a connected dominating set $D$ that induces a subgraph 
that is either $P_7$-free or isomorphic to a $P_7$.
If $G|D$ is isomorphic to a $P_7$, then $D$ is the desired set. Otherwise we apply \autoref{thm:Pt} on $G|D$ to conclude that
$G|D$ has a connected dominating set $D'$ that induces a subgraph that is either $P_5$-free or isomorphic to a $P_5$.
If $G|D'$ is isomorphic to a $P_5$, then $D'$ is the desired set. Otherwise $G|D'$ is $P_5$-free.
We again apply \autoref{thm:Pt} on $G|D'$: $G|D'$ has a connected dominating set $D''$ that \
induces a subgraph that is either $P_3$-free or isomorphic to a $P_3$.
Then $D''\cup N(D'')\cup N(N(D''))$ dominates $G$.
Since $G$ is triangle-free, if $G|D''$ is $P_3$-free, then $D''$ is a clique of size at most 2.
It follows that $|D''|\le 3$ and so $D''$ is the desired set. 
\end{inproof}

Let $S$ be the connected set guaranteed by \autoref{seed}. Define $X=N(S)$, $Y=N(N(S))\setminus S$ and $Z=V(G)\setminus (X\cup Y\cup Z)$.
Then $(S,X,Y,Z)$ is a partition of $V(G)$,  $S$ dominates $X$, $X$ dominates $Y$ , and
there is no edge between $S$ and $Y\cup Z$ or between $X$ and $Z$. Moreover it follows from \autoref{seed} that $Y$ dominates $Z$.
Such a quadruple $(S,X,Y,Z)$ is called a {\em layer structure} of $G$.
We write $\mathcal{P}=(S,X,Y,Z)$. The set $S$ is called the {\em seed} for $\mathcal{P}$.

\vspace*{0.3cm}
\noindent {\bf Phase II.  Obtaining a canonical $C_k$-list assignment via updates.}
\vspace*{0.3cm}

We now branch on every possible coloring of $S$, respecting the lists $L$.
Since $|S|\le 7$, there are at most $7^{k}$ such colorings of $G|S$.
Note that $7^k$ is a constant since $k$ is a fixed number. 
To prove the theorem, therefore,  it suffices to determine whether a given coloring $f:S\rightarrow [k]$ 
can be extended to a $C_k$-coloring of $(G,L)$ in polynomial time.
In the following, we fix such a coloring $f:S\rightarrow [k]$, and therefore, we are dealing with an instance $(G,L')$ where 
\[L'(v)=\left\{
\begin{array}{ll}
L(v) & \mbox{if $v\notin S$,} \\
\{f(v)\} & \mbox{if $v\in S$.}
\end{array}
\right.\]

We further partition the sets $S$, $X$ and $Y$ as follows.
For $1\le i\le k$, let
\begin{align*}
& S_i=\{s\in S:L(s)=\{i\}\},\\
& X_i=\{x\in X\setminus (\bigcup_{j=1}^{i-1}X_j):N(x)\cap S_i\neq \emptyset.\},\\
& Y_i=\{y\in Y\setminus (\bigcup_{j=1}^{i-1}Y_j):N(y)\cap X_i\neq \emptyset.\}.
\end{align*}
Clearly, $(X_1,X_2,\ldots,X_k)$ is a partition of $X$ and $(Y_1,Y_2,\ldots,Y_k)$ is a partition of $Y$.

%
%
%

We now perform the following updates for all $1\le i\le k$ in the following order.

\begin{itemize}
  \item For every edge $sx$ with $s\in S_i$ and $x\in X_i$, we update $x$ from $s$.
  \item For every edge $xy$ with $x\in X_i$ and $y\in Y_i$, we update $y$ from $x$.
\end{itemize}

We continue to denote the resulting $C_k$-list assignment by $L'$. Then $|L'(s)|=1$ for every $s\in S$, 
$L'(x)\subseteq  \{i-1,i+1\}$ for every $x\in X_i$  and $L'(y)\subseteq  \{i,i-2,i+2\}$ for every $y\in Y_i$. 
We call such a $C_k$-list assignment $L'$ {\em canonical} for $\mathcal{P}=(S,\bigcup_{i=1}^{k}X_i,\bigcup_{i=1}^{k}Y_i,Z)$.

\begin{claim}\label{clm:Xistable}
If $X_i$ is not a stable set, then $(G,L')$ is not $C_k$-colorable.
\end{claim}

\begin{inproof}
  Suppose that $X_i$ contains an edge $uv$. In every $C_k$-coloring $g$ of $(G, L')$, there is a $j \in [k]$ such that $\{g(u),g(v)\}=\{j,j+1\}$. Since $L'(u),L'(v)\subseteq \{i-1,i+1\}$, it follows  that no such $C_k$-coloring exists. 
\end{inproof}

Note that one can determine in $O(n^2)$ time if there exists an $X_i$ that is not stable. If so, we stop and correctly determine
that $(G,L')$ is not $C_k$-colorable by \autoref{clm:Xistable}. 
Otherwise,  we may assume that $X_i$ is stable for all $1\le i\le k$ from now on.

\vspace*{0.3cm}
\noindent {\bf Phase III.  Eliminating bad paths via branching} ($O(n^{12k})$ branches).
\vspace*{0.3cm}

In this phase, we shall reduce the instance $(G,L')$ to an equivalent set of polynomially many subinstances
so that every subinstance has no bad paths, which we define now.

\noindent {\bf Definition (Bad path).}
An induced path $a-b-c$ is a \emph{bad path} in $\mathcal{P}=(S,X,Y,Z)=(S,\bigcup_{i=1}^{k}X_i,\bigcup_{i=1}^{k}Y_i,Z)$ 
if for some $i\in [k]$, $a\in Y_i$, $b,c\in (Y\cup Z)\setminus  Y_i$ 
and $\{b,c\}$ is anticomplete to $X_i$. We call $a$ the {\em starter} of $a-b-c$.
Let $\mathcal{P}_i$ be the set of all bad paths with starters in $Y_i$. Note that $|\mathcal{P}_i|=O(n^3)$.

\begin{center}
		\begin{tikzpicture}
			\node at (2.5,0) {$ S $};
			\draw (2.5,0) ellipse (1 and 2);
			\draw (2.5,1.2) ellipse (0.6 and 0.4);
			\node at (2.5,1.2) {\small $ S_1 $};
			\draw[gray, thick] (3.1,1.2) -- (4.9,1.2);
			\node at (5.5,0) {$ X $};
			\draw (5.5,0) ellipse (1 and 2);
			\draw (5.5,1.2) ellipse (0.6 and 0.4);
			\node at (5.5,1.2) {\small $ X_1 $};
			
			\draw[gray, thick] (6.1,1.2) -- (7.9,1.2);
			\node at (8.5,0) {$ Y $};
			\draw (8.5,0) ellipse (1 and 2);
			\draw (8.5,1.2) ellipse (0.6 and 0.4);
			\node at (8.5,1.2) {\small $ Y_1 $};
			\filldraw[black] (8.8,1.4) circle (2pt) node[anchor=north]{a};
			\node at (11.5,0) {$ Z $};
			\draw (11.5,0) ellipse (1 and 2);
			\filldraw[black] (11.5,1.4) circle (2pt) node[anchor=west]{b};
			\filldraw[black] (11.7,0.5) circle (2pt) node[anchor=west]{c};
			\draw[gray, thick] (8.8,1.4) -- (11.5,1.4) -- (11.7,0.5);
		\end{tikzpicture}

\captionof{figure}{\small An illustration of the layer structure, where the induced path $a-b-c$ is an example of a bad path.}
\end{center}

\noindent {\bf Definition (Depth).}
A vertex $v\in Y_i$ is of \emph{depth $\ell$} to the seed $S$ if for every $x\in N(v)\cap X_i$, 
there exists an induced  path $v-x-P$ of length $\ell$ such that $V(P)\subseteq S$.

Observe that every vertex in $Y$ is of depth at least 3 to $S$ (because we may assume that $|S|\ge 2$ and so no vertex in $X$ is complete to $S$
since $G$ is triangle-free),  and that the starter of a bad path is of depth at most 7 to $S$ since $G$ is $P_9$-free.

Note that for any $C_k$-coloring of $(G,L')$  (if one exists), either there exists a bad path in $\mathcal{P}_i$ whose starter
is colored with a color in $\{i-2,i+2\}$ or the starters of all bad paths in $\mathcal{P}_i$ are colored with color $i$.
This leads to the following branching scheme.

\medskip
\noindent {\bf Branching.} (List change only.)

\begin{itemize}
\item ($2^k=O(1)$ branches.)
  
For every subset $I\subseteq [k]$, we have a branch $B_I$ intended to find possible colorings such that there exists a bad path in $\mathcal{P}_i$ whose starter is colored with a color in $\{i-2,i+2\}$ if $i\in I$, and all starters of bad paths in $\mathcal{P}_i$ are colored with color $i$ if $i\notin I$. Clearly, $(G,L')$ is $C_k$-colorable if and only if at least one of the $B_I$ is a yes-instance. In the following, we fix a branch $B_I$.

\item ($O(2^kn^{3k})=O(n^{3k})$ branches.)
  
We further branch to obtain a set of size $O(n^{3k})$ of subinstances within $B_I$ by guessing, for each $i \in I$, a bad path 
in $\mathcal{P}_i$, and guessing the color of its starter from $\{i-2,i+2\}$. The union over all branches $B_I$ of these subinstances is equivalent to $(G, L')$. 
\end{itemize}
Specifically, for each element $(a_i, b_i, c_i)_{i \in I}$
in $\Pi_{i \in I} \mathcal{P}_i$, we have one branch
where we set $L''(a_i):=L'(a_i)\cap \{i-2,i+2\}$ for every $i\in I$, and we set
$L''(a):=L'(a)\cap \{i\}$ for every starter $a$ of a bad path in $\mathcal{P}_i$ for every $i\notin I$.  
We denote the resulting $C_k$-list assignment by $L''$.
For each such branch and for every element $(q_i)_{i \in I}$ in $\Pi_{i \in I} L''(a_i)$,
we have one branch where $L''(a_i):=\{q_i\}$ for all $i \in I$. It follows that for all $i \in I$ and $x \in X_i \cap N(a_i)$, the only possible color for $x$ is $i+1$ if $q=i+2$ and $i-1$ if $q=i-2$, and so we set $L''(x) = \{(q_i + i)/2\}$. 
Since $L''(a_i)\subseteq \{i-2,i+2\}$ for all $i \in I$, it follows that there are $2^{|I|}\le 2^k$ branches for each set of bad paths we guess, which gives $O(2^kn^{3k})$ branches in the second step of branching.
Let us fix one such branch and denote the resulting instance by $(G,L'')$.
\begin{itemize}
\item ($O(k^{3k})=O(1)$ branches.)
  
We let $I^*$ be the subset of $[k] \setminus I$ of indices $i$ such that $\mathcal{P}_i$ contains a bad path. For each $i \in I^*$, we choose a bad path $a_i-b_i-c_i$ in $\mathcal{P}_i$ such that $|N(a_i)\cap X_i|$ is minimum, 
where the minimum is taken over all bad paths in $\mathcal{P}_i$. Choose a vertex $x_i\in N(a_i)\cap X_i$ for each $i \in I^*$.
Let 
\[
	Q=\bigcup_{i\in I}\{b_i,c_i\}\cup \bigcup_{i \in I^*}\{b_i,c_i,x_i\},
\]
where for $i\in I$, $b_i,c_i$ are two vertices on the bad path we guessed in the previous bullet. 
We branch on every possible coloring of $Q$, respecting the lists $L$. Since $|Q|\le 3k$, the number of branches is at most $k^{3k}$. In the following, we fix a coloring $g$ of $Q$ and denote the resulting subinstance by $(G,L''')$, where
\[L'''(v)=\left\{
\begin{array}{ll}
L''(v) & \mbox{if $v\notin Q$}, \\
\{g(v)\} & \mbox{if $v\in Q$}.
\end{array}
\right.\]
\end{itemize}

\noindent {\bf Obtaining a new layer structure with a canonical $C_k$-list assignment.}
\vspace*{0.3cm}

 We now deal with $(G,L''')$.
Define
\[
	A=\bigcup_{i\in I}\left(\left(N(a_i)\cap X_i\right)\cup \{a_i,b_i,c_i\}\right)\cup \bigcup_{i\in I^*}\{x_i,a_i,b_i,c_i\},
      \]
 and note that in $L'''$, every vertex in $A$ has a list of size at most 1. 
We update all vertices of $G$ from all vertices in $A$ and continue to denote the resulting $C_k$-list assignment by $L'''$.
We now obtain a new partition $\mathcal{P}'=(S',X',Y',Z')$ of $G$ as follows.

\begin{itemize}
\item Let $S'=S\cup A$. 

\item For each $1\le j\le k$, let $K_j:=\emptyset$. For each vertex $v\in Y\cup Z$, 
if $v$ has a neighbor in $S'$, let $j$ be the smallest integer in $[k]$ such that there exists a vertex $s\in N(v)\cap S'$
with $L(s)=\{j\}$, and add $v$ to $K_j$. For each $1\le j\le k$, let $X'_j=(X_j\cup K_j)\setminus A$.
Let $X'=\bigcup_{i=1}^{k}X'_i$.

\item For $1\le i\le k$, let $Y'_i$ be the set of vertices in 
$V(G)\setminus (S'\cup X'\cup (\bigcup_{j< i}Y'_j))$ that have a neighbor in $X'_i$.
Let $Y'=\bigcup_{i=1}^{k}Y'_i$.

\item Let $Z'=V(G)\setminus (S'\cup X'\cup Y')$.
\end{itemize}

\begin{claim}
The new partition $\mathcal{P}'=(S',X',Y',Z')$ is a layer structure of $G$ and $L'''$ is a canonical $C_k$-list assignment for $\mathcal{P}'$.
\end{claim}

\begin{inproof}
From the definition of $(S',X',Y',Z')$, it follows that $S'$ dominates $X'$,  $X'$ dominates $Y'$ and $Y'$ dominates $Z'$.
There are no edges between $S'$ and $Y'\cup Z'$ and no edges between $X'$ and $Z'$. 
Moreover, $S'$ is connected by the definition of $A$ and the fact that $S$ is connected.
So  $\mathcal{P}'=(S',X',Y',Z')$ is a layer structure. 
Note that $|L'''(s)|=1$ for every $s\in S'$,
$L'''(x)\subseteq \{i-1,i+1\}$ for every $x\in X'_i$ and $L'''(y)\subseteq \{i,i-2,i+2\}$ for every $y\in Y'_i$.
So $L'''$ is canonical for $\mathcal{P}'=\left(S',\bigcup_{i=1}^{k}X'_i,\bigcup_{i=1}^{k}Y'_i,Z'\right)$.
\end{inproof}

\begin{claim}
The following hold for all $i\in[k]$.
\begin{enumerate}[label=(\arabic*)]
\item $ X'_i\setminus X_i\subseteq Y\cup Z $.\label{item:1}
\item If a vertex in $ Y'\cup Z'$ is anticomplete to $X'_i$, then it is anticomplete to $X_i$.\label{item:2}
\item $Y'_i\setminus Y_i$ is anticomplete to $X_i$.\label{item:3}
\end{enumerate}
\end{claim}

\begin{inproof}
By construction, $X'_i\setminus X_i\subseteq K_i\subseteq Y\cup Z$ and so \ref{item:1} follows.

Let $v\in  Y'\cup Z'$ be anticomplete to $X'_i$.  Since $v\notin S'\cup X'$, it follows that $v$ is anticomplete to $A$.
Note that $ X_i\setminus X'_i\subseteq A$. So \ref{item:2} follows from the assumption that $v$ is anticomplete to $X'_i$.

Suppose for a contradiction to \ref{item:3} that there exists $y\in Y'_i\setminus Y_i$ that has a neighbor in $X_i$.
Since $y\in Y'_i$ and $Z$ is anticomplete to $X$, it follows that $y\in Y_j$ for some $j\neq i$. From the definition of the sets $Y_1,\ldots,Y_k$,
it follows that $y\in Y_j$ for some $j<i$. Let $x\in X_j$ be a neighbor of $y$. It follows that $x\notin X'_j$, for otherwise $y$ would be
in $Y'_k$ for some $k\le j$, which contradicts the assumption that $y\in Y'_i$. 
So $x\in A$, but then $y \in X'$, a contradiction. So \ref{item:3} follows.
\end{inproof}

The following claim is the key to our branching algorithm.

\begin{claim}\label{clm:depth}
Let $a$ be a starter of a bad path in $\mathcal{P}'$. If the depth of the starter of any bad path in $\mathcal{P}$ is at least $\ell$, then the depth of $a$ in $\mathcal{P}'$ is at least $\ell+1$.
\end{claim}

\begin{inproof}
Let $a'-b'-c'$ be a bad path in $\mathcal{P}'$ with $a'\in Y'_i$. We consider the following two cases.
\vspace*{0.3cm}

\noindent {\bf Case 1:} 
$a'\in Y_i\cap Y_i'$.
\vspace*{0.3cm}

Then $\emptyset\neq N(a')\cap X_i\subseteq X'_i$. 
By \ref{item:2}, $\{b',c'\}$ is anticomplete to $X_i$ and so $a'-b'-c'$ is also a bad path in $\mathcal{P}=(S,X,Y,Z)$. 
This implies that $\mathcal{P}_i\neq \emptyset$. Therefore, there exist $a,b,c,x\in S'$ such that 
$a-b-c$ is a bad path in $\mathcal{P}$ with $a\in Y_i$ and $x\in N(a)\cap X_i$.

We first claim that it is possible to pick a vertex $x'\in N(a')\cap X_i$ that is not adjacent to $a$.
Recall that the branch we consider corresponds to a set $I \subseteq [k]$.
If $i\in I$, then all vertices in $N(a)\cap X_i$ are in $A$ and hence are now in $S'$. 
So every vertex in $N(a')\cap X_i$ is not adjacent to $a$, and our claim holds. If $i \not\in I$, then $i \in I^*$, and so $a=a_i$. By the choice of $a_i$, it follows that $|N(a)\cap X_i|\leq |N(a')\cap X_i|$.  
Since $a'\in Y_i'$, it follows that $a'$ is not adjacent to $x$. Therefore, there exists a vertex $x'\in N(a')\cap X_i$ 
such that $x'$ is not adjacent to $a$.

Note that $x$ and $x'$ are not adjacent by \autoref{clm:Xistable}.
Moreover, $x'$ is anticomplete to $\{b',c',b,c\}$ by the definition of bad path. 
Let $P'$ be the shortest path from $x$ to $x'$ with internal vertices contained in $S$. 
Note that $P'$ exists since $S$ is connected.
Then $P'$ is an induced path. Since $V(P')\setminus \{x,x'\}\subseteq S$,  it follows that $V(P')\setminus \{x,x'\}$ is anticomplete
to $\{a,b,c,a',b',c'\}$. Therefore, $c-b-a-x-P'-x'-a'-b'-c'$ is an induced path of order at least 9, a contradiction.
\vspace*{0.3cm}

\noindent {\bf Case 2:} 
$a'\in Y'_i\setminus Y_i$.
\vspace*{0.3cm}

It follows from \ref{item:3} that $N(a')\cap X'_i\subseteq X_i'\setminus X_i$. 
Pick a vertex $x'\in N(a')\cap X'_i$. Since $x\in X'_i\setminus X_i$, $x'$ has a neighbor $s'\in S'$ by the definition of $X'_i$. 
By \ref{item:1}, $x'\in  Y\cup Z$ and so $s'\in S'\setminus S=A$.
This implies that there exists an index $j\in I$ such that $x'$ is not anticomplete to $Q=\{x_j,a_j,b_j,c_j\}$
where $x_j\in N(a_j)\cap X_j$. Let $a_j-x_j-P$ be an induced path of length $\ell$ with $V(P)\subseteq S$.
Note that $x'\in  Y\cup Z$ is anticomplete to $V(P)$. Let $x'-P''-x_j$ be the shortest path from $x'$ to $x_j$ 
such that $V(P'')\subseteq Q$. Since  $a'$ is anticomplete to $\{x\}\cup V(P)\cup V(P'')\subseteq S'$,
it follows that $a'-x'-P''-x_j-P$ is an induced path of length at least $\ell+1$. This proves the claim.
\end{inproof}

Therefore, we have obtained an equivalent set of subinstances of size $O(n^{3k})$.
For each such subinstance, the minimum depth of the starter of a bad path has increased by at least 1 compared to $\mathcal{P}$ due to \autoref{clm:depth}. Note that the depth of any starter of a bad path in $\mathcal{P}$ is at least 3.
Moreover, since $G$ is $P_9$-free, the depth of any starter of a bad path is at most 7.
By branching 4 times, therefore, we obtain an equivalent set of $O(n^{12k})$ subinstances such that each subinstance has no bad paths.

\vspace*{0.3cm}
\noindent {\bf Phase IV.  Reducing the list size of vertices in $Z$.}

\vspace*{0.3cm}
Let us now fix an instance $(G,L)$ where  $\mathcal{P}=(S,X,Y,Z)$ is a layer structure with no bad paths
and $L$ is canonical for $\mathcal{P}$. The goal of this phase is to reduce the list size of vertices in $Z$. We start with a property of $Z$. 

\begin{claim}\label{clm:Zstable}
The set $Z$ is stable and each $z\in Z$ has neighbors in at most one of $\{Y_1,Y_2,\ldots,Y_k\}$.
\end{claim}

\begin{inproof}
Suppose by contradiction that $Z$ has an edge $z_1z_2$. Let $y\in Y$ be a neighbor of $z_1$.
Since $G$ is triangle-free, $z_2$ is not adjacent to $y$. Then $y-z_1-z_2$ is a bad path, a contradiction.
So $Z$ is stable.
Suppose that $z\in Z$ has a neighbor $y_i\in  Y_i$ and $y_j\in Y_j$ for $i\neq j$. We may assume that $i<j$.
Then by the definition of $Y_1,Y_2,\ldots,Y_k$,  it follows that $y_j$ is anticomplete to $X_i$. So $y_i-z-y_j$
is a bad path, a contradiction.
\end{inproof}

 For $i \in [k]$, we let $Z_i = N(Y_i) \cap Z$. By \autoref{clm:Zstable}, it follows that $Z_1, \dots, Z_k$ is a partition of $Z$. Now we are ready to reduce the list size of vertices in $Z$, depending on which subset it belongs to.
\begin{claim}\label{clm:Z}
There is an equivalent instance $(G, L')$ for $(G,L)$ such that $L'(z)$ is a subset of $ \{i+1,i-1\},\{i+1,i-3\},\{i-1,i+3\}$ or $\{i+3,i-3\}$ for all $z \in Z_i$. 
\end{claim}

\begin{inproof}
For $z \in Z_i$, we define $c_1(z) = \{i-1\}$ if $i-1 \in L(z)$, and $c_1(z) = \{i-3\} \cap L(z)$, otherwise; we define $c_2(z) = \{i+1\}$ if $i+1 \in L(z)$, and $c_2(z) = \{i+3\} \cap L(z)$, otherwise. Now let $L'(z) = c_1(z) \cup c_2(z)$ for all $z \in Z$. It follows that $L'(z)$ is a subset of $ \{i+1,i-1\},\{i+1,i-3\},\{i-1,i+3\}$ or $\{i+3,i-3\}$ for all $z \in Z_i$.

  Since $(G, L')$ is a subinstance of $(G, L)$, it follows that if $(G, L')$ has a $C_k$-coloring, then so does $(G, L)$. Now suppose that $(G, L)$ has a $C_k$-coloring $c$, and choose $c$ such that $c(z) \in L'(z)$ for as many $z \in Z$ as possible. If $c(z) \in L'(z)$ for all $z \in Z$, then $c$ is a $C_k$-coloring of $(G, L')$, and so $(G, L')$ is equivalent to $(G, L)$ and the claim follows.

  Now suppose for a contradiction that there is a vertex $z \in Z$ such that $c(z) \not\in L'(z)$. Since every vertex $y \in Y_i$ satisfies $L(y) \subseteq \{i, i+2, i-2\}$,  $c(z) \in \{i+1, i-1, i+3, i-3\}$. 
   If $c(z)\in \{i+1,i-1\}$, since $c(z)\in L(z)$, by the definition of $L'(z)$ it follows that $c(z)\in L'(z)$, a contradiction. So $c(z) \in \{i-3, i+3\}$. By symmetry, we may assume that $c(z) = i+3$. Since $c(z) \not\in L'(z)$, it follows that $c_2(z) = i+1$, and therefore $i+1 \in L(z)$. Let $y \in N(z)$. By \autoref{clm:Zstable}, it follows that $y \in Y_i$, and consequently, $L(y) \subseteq \{i,i-2,i+2\}$. Since $c(z) = i+3$, and $c$ is a $C_k$-coloring, it follows that $c(y) = i+2$ for all $y \in N(z)$. Now define $c'$ by letting $c'(z) = i+1$, and $c'(v) = c(v)$ for all $v \neq z$. It follows that $c'$ is a $C_k$-coloring of $(G, L)$, contrary to the choice of $c$. This is a contradiction, and the claim follows. 
\end{inproof}

We now modify the lists of vertices in $Z$ as in the $C_k$-list assignment $L'$ of \autoref{clm:Z}, and we continue to denote by the resulting list $L$. 

\medskip
\noindent {\bf Phase V.  Reducing the list size of vertices in $Y$.}

\medskip

Now we have obtained a layer structure with a canonical $C_k$-list assignment such that no bad path exists and the list size of vertices in $Z$ has been reduced. In this phase, we start with updating vertices in $G|S\cup X\cup Y$ and rearranging the vertices based on their lists; then we further reduce the list size based on the structure of components in $Y'\cup Z'$ so that we can apply \autoref{thm:2SAT}.

We first apply \autoref{lem:reduced list} on $G|S\cup X\cup Y$ to obtain a $C_k$-list assignment $L'$ which is reduced on $G|S\cup X\cup Y$.
Then $(G,L')$ is an equivalent subinstance of $(G,L)$.

If $L'(v)=\emptyset$ for some $v\in V(G)$, we stop and claim that $(G,L)$ is not $C_k$-colorable. Otherwise define
\begin{align*}
& S'=\{v\in S\cup X\cup Y:|L'(v)|=1\},\\
& X'_i=\{v\in (X\cup Y)\setminus S':L'(v)\subseteq \{i-1,i+1\}\}, 1\le i\le k,\\
& Y'_i=\{v\in Y\setminus (S'\cup X'\cup \bigcup_{j<i}Y'_j):L'(v)\subseteq \{i,i-2,i+2\}\}, 1\le i\le k,\\
& Z'=Z,\\
& X'=\bigcup_{i=1}^{k}X'_i,\\
& Y'=\bigcup_{i=1}^{k}Y'_i.
\end{align*}

Recall that we have modified the list of $z\in Z$ according to \autoref{clm:Z}. It follows that $|L'(v)|\leq 3$ for every $v\in Y'$ and $|L'(v)|\leq 2$ for every $v\in G\setminus Y'$. 

Also recall that $\mathcal{P}=(S,X,Y,Z)$ is a layer structure of $G$ that is given at the beginning of Phase IV, and we have defined $ S_i=\{s\in S:L(s)=\{i\}\},
     X_i=\{x\in X\setminus (\bigcup_{j=1}^{i-1}X_j):N(x)\cap S_i\neq \emptyset.\},
    Y_i=\{y\in Y\setminus (\bigcup_{j=1}^{i-1}Y_j):N(y)\cap X_i\neq \emptyset.\}
$ for each $i\in [k]$.

Next, we prove a few properties for $S'$, $X'$, $Y'$ and $Z'$, which shows how those sets related to the previous layer structure and give properties for components in $Y'\cup Z'$ so that we can further reduce the lists.

\begin{claim}\label{clm:update everything}
The following hold for $S'$, $X'$, $Y'$ and $Z'$.
\begin{enumerate}[label=(\arabic*)]
\item $V(G)=S'\cup X'\cup Y'\cup Z'$. \label{item:4}
\item For every $i\in [k]$ and $y\in Y'_i$, we have $N(y)\cap (S'\cup X')\subseteq X'_i$.\label{item:5}
\item For every $i\in [k]$, we have $X_i\subseteq X'_i\cup S'$, $Y'_i\subseteq Y_i$ and $Y_i\subseteq Y'_i\cup S'\cup X'$.\label{item:6}
\item There does not exist an induced path $a-b-c$ such that $a\in Y'_i$, $b,c\in (Y'\cup Z')\setminus Y'_i$.\label{item:6.5}
\item Let $C$ be a component in $Y'\cup Z'$ such that $V(C)\cap Y'\neq \emptyset$, then one of the followings holds:\label{item:7}
\begin{enumerate}[label=(\alph*)]
	\item $V(C)\subseteq Y'_i\cup Z'$ for some $i\in [k]$; or
		\item $V(C)\subseteq Y'_i\cup Y'_j$ for some $i,j\in [k]$ and every edge in $C$ has one end in $Y_i$ and the other in $Y_j$;
\end{enumerate}
\end{enumerate}
\end{claim}

\begin{inproof}
If $v\in  S$, then $v\in S'$. If $v\in X$, then $s\in S'\cup X'$. If $v\in Y$, then $v\in S'\cup X'\cup Y'$.
This proves \ref{item:4}.


Let $y\in Y'_i$. Then $L'(y)\subseteq \{i-2,i,i+2\}$. Since $y\notin S'\cup X'$, it follows that $L'(y)=\{i-2,i+2\}$ or 
$L'(y)=\{i-2,i,i+2\}$. Since $L'$ is reduced on $G|S'\cup X'\cup Y'$, $N(y)\cap S'=\emptyset$.
Pick an arbitrary vertex $x\in N(y)\cap X'_j$ for some $j\in [k]$.  Then $L'(x)=\{j+1,j-1\}$. We show that $i=j$.
Since $L'$ is reduced on $G|S'\cup X'\cup Y'$, we cannot effectively update $y$ from $x$. It follows that $L'(y)\subseteq \{j-2,j,j+2\}$. 
If $L'(y)=\{i-2,i,i+2\}$, then either $i=j$ or $i=j\pm 2$ and $k=6$. 
If $L'(y)=\{i-2,i+2\}$, then either $i=j$, $i=j\pm 2$ and $k=6$ or $i=j\pm 4$ and $k=8$.
Since $k\neq 6,8$, it follows that $i=j$.
This proves \ref{item:5}.


Let $x\in X_i$. Then $L(x)\subseteq \{i-1,i+1\}$. Since update cannot make the list larger, it follows that $L'(x)\subseteq L(x)$
and so $x\in X'_i\cup S'$. Let $y\in Y'_i$. Then $L'(y)\subseteq \{i,i-2,i+2\}$. 
Since $y\notin S'\cup X'$, it follows that $L'(y)=\{i-2,i+2\}$ or  $L'(y)=\{i-2,i,i+2\}$.
 If $y\notin Y_i$, then $y\in Y_j$ for some $j\neq i$.
So $L(y)\subseteq \{j,j-2,j+2\}$. Then $L'(y)\subseteq \{j-2,j,j+2\}$. 
If $L'(y)=\{i-2,i,i+2\}$, then either $i=j$ or $i-j=\pm 2$ and $k=6$. 
If $L'(y)=\{i-2,i+2\}$, then either $i=j$ or $i-j=\pm 2$ and $k=6$ or $i-j=\pm 4$ and $k=8$.
All cases lead to a contradiction, so $y\in Y_i$ and $Y'_i\subseteq Y_i$.
Let $y\in Y_i$, then by construction, $y\in S'\cup X'\cup Y'$. If $y\in Y'_j$ for some $j\neq i$, then $y\in Y_j$, contrary to $y\in Y_i$. So $y\notin Y'_j$ with $j\neq i$ and it follows that $Y_i\subseteq Y'_i\cup S'\cup X'$.
This proves \ref{item:6}.

To prove \ref{item:6.5}, it is sufficient to show that such a path $a-b-c$ is a bad path in $\mathcal{P}$. By \ref{item:6} and the construction,   $a\in Y_i$, $b,c\in Y\cup Z\setminus Y_i$. If $w\in\{b,c\}$ is adjacent to $x\in X_i$, then $w\in Y\setminus Y_i$. We may assume $w\in Y_j$ for $j\neq i$, then by \ref{item:6} $w\in Y'_j$ and by \ref{item:5} $x\in X'_j$, a contradiction to $x\in X_i\subseteq X_i'\cup S'$. It follows that $\{b,c\}$ is anticomplete to $X_i$ and $a-b-c$ is a bad path in $\mathcal{P}$. This proves \ref{item:6.5}.

Let $C$ be a component in $Y'\cup Z'$ such that $V(C)\cap Y'\neq \emptyset$. By \autoref{clm:Zstable}, $Z'=Z$ is stable and for every $z\in Z$, $N(z)\subseteq Y_i$ for some $i$. Hence for every $z\in C$, $N(z)\cap V(C)\subseteq Y'_i$ for some $i$. Assume that \ref{item:7}.(a)  does not hold, then there exists an edge $uv\in C$ such that $u\in Y'_i$, $v\in Y'_j$, $i,j\in [k]$ with $i\neq j$. Suppose for a contradiction there exists $w\in C$ with $w\in Y'_k\cup Z$ for some $k\neq i,j$. Let $u-v-p_1-p_2-\dots-p_n=w$ be the shortest path from $\{u,v\}$ to $w$ in $C$. Since $u\in Y'_i$ and $v\in Y'_j$, it follows from \ref{item:6.5} that $p_1\in Y'_i$, and then $p_2\in Y'_j$, $p_3\in Y'_i$, and so on. Inductively, it follows that $w=p_n\in Y'_i\cup Y'_j$, a contradiction. So $V(C)\subseteq Y'_i\cup Y'_j$. By \ref{item:6.5} and the triangle-freeness of $G$, every edge in $C$ has one end in $Y'_i$ and the other in $Y'_j$.
This proves \ref{item:7}.
\end{inproof}

We construct a $C_k$-list assignment $L''$ as follows: for every component $C$ in $Y'\cup Z'$ such that $V(C)\cap Y'\neq \emptyset$, 
\begin{itemize}
	\item If $V(C)\subseteq Y'_i$ for some $i\in [k]$ and $|V(C)|\geq 2$, for every $v\in V(C)$, set $L''(v)= \{i+2,i-2\}\cap L'(v)$ if $k=5$ and set $L''(v)=\emptyset$ otherwise;
	\item If  $V(C)\subseteq Y'_i$ for some $i\in [k]$ and $|V(C)|=1$, for $v\in V(C)$, set $L''(v)=L'(v)$ if $|L'(v)|<3$ and $L''(v)=\{i\}$ if $|L'(v)|=3$;
	\item If $V(C)\subseteq Y'_i\cup Z'$ for some $i\in [k]$ and there exists an edge in $G|(V(C)\cap Y'_i)$,  for every $v\in V(C)\cap Y'_i$, set $L''(v)= \{i+2,i-2\}\cap L'(v)$ if $k=5$ and set $L''(v)=\emptyset$ otherwise;
	\item If $V(C)\subseteq Y'_i\cup Y'_{i+1}$ for some $i\in [k]$,  for every $v\in V(C)\cap Y'_i$ with $|L'(v)|=3$ set $L''(v)=  L'(v)\setminus \{i-2\}$ and  for every $v\in V(C)\cap Y'_{i+1}$ with $|L'(v)|=3$ set $L''(v)=  L'(v)\setminus \{i+3\}$; 
\end{itemize}
And set $L''(v)=L'(v)$ for every other vertex in $G$. It is clear that $|L''(v)|\leq 3$ for every $v\in G$ and $|L''(v)|\leq 2$ for every $v\in G\setminus Y'$. Next we prove that $ (G,L'') $ is an equivalent subinstance of $(G,L)$ and we can apply \autoref{thm:2SAT} on $(G,L'')$.

\begin{claim}\label{clm:reduce  Y}
	The following holds for $(G,L'')$:
	\begin{enumerate}[label=(\arabic*)]
		\item $ (G,L'') $ is an equivalent subinstance of $(G,L)$. \label{item:8}
		\item Let $A=\{v\in V(G) ~:~ |L''(v)|=3 \}$, then $A$ is a stable set and \label{item:9}
		\begin{itemize}
			\item For every $v\in V(G)\setminus A$, $|L(v)|\le 2$.
			\item For every $v\in A$, $L(v)=\{i,i-2,i+2\}$ for some $i\in [k]$ and for every $u\in N(v)$, $L(u)$ is a subset of  $ \{i+1,i-1\},\{i+1,i-3\},\{i-1,i+3\}$ or $\{i+3,i-3\}$. 
		\end{itemize} 
			\end{enumerate}
\end{claim}
\begin{inproof}
	Recall that for $v\in Y'_i$, if $|L'(v)|=3$, then $L'(v)=\{i,i-2,i+2\}$. It follows from the construction of $L''$ that $ (G,L'') $ is an subinstance of $(G,L')$ and therefore if $(G, L'')$ has a $C_k$-coloring, then so does $(G, L')$. Now suppose that $(G, L')$ has a $C_k$-coloring $c$, and choose $c$ such that $c(v)\in L''(v)$ for as many $v\in G$ as possible.  Suppose for a contradiction that there exists $v\in G$ such that $c(v)\notin L''(v)$, then $v\in C$, where $C$ is a component $C$ in $Y'\cup Z'$ such that $V(C)\cap Y'\neq \emptyset$ and one of the following cases hold:
	\begin{itemize}
		\item Case 1: $V(C)\subseteq Y'_i$ for some $i\in [k]$ and $|V(C)|\geq 2$. Since there exists an edge in $G|(V(C)\cap Y'_i)$ and $L'(u)\subseteq \{i,i-2,i+2\}$ for every $u\in Y'_i$ , it follows that $k=5$,  $L''(v)= \{i+2,i-2\}\cap L'(v)$ and $c(v)\in \{i-2,i+2\}$, a contradiction to $c(v)\notin L''(v)$.
		\item Case 2:  $V(C)\subseteq Y'_i$ for some $i\in [k]$ and $|V(C)|=1$. Then $L'(v)=\{i,i+2,i-2\}$ and $L''(v)=\{i\}$. By \autoref{clm:update everything}, $N(v)\subseteq X'_i$. It follows that for every $u\in N(v)$, $L(u)\subseteq \{i-1,i+1\}$. Now define $c'$ by letting $c'(v)=i$ and $c'(u)=c(u)$ for every $u\neq v$. It follows that $c'$ is a $C_k$-coloring of $(G,L')$, contrary to the choice of $c$.
		\item  Case 3: $V(C)\subseteq Y'_i\cup Z'$ for some $i\in [k]$ and there exists an edge in $G|(V(C)\cap Y'_i)$. Similarly to Case 1, it follows that $k=5$,  $L''(v)= \{i+2,i-2\}\cap L'(v)$ and $c(v)\in \{i-2,i+2\}$, a contradiction to $c(v)\notin L''(v)$.
		\item Case 4:  $V(C)\subseteq Y'_i\cup Y'_{i+1}$ for some $i\in [k]$. We may assume $v\in Y'_i$, then $L'(v)=\{i,i+2,i-2\}$, $L''(v)=\{i,i+2\}$ and $c(v)=i-2$. It follows from \autoref{clm:update everything} that $N(v)\subseteq Y_{i+1}\cup X_i$. Let $N=\bigcup_{u\in N(v)}\{c(u)\} $, then $N\subseteq \bigcup_{u\in Y_{i+1}\cup X_i}L(u)\subseteq \{i-1,i+1,i+3\}$. Since $c(v)=i-2$, $N=\{i-1\}$.  Now define $c'$ by letting $c'(v)=i$ and $c'(u)=c(u)$ for every $u\neq v$. It follows that $c'$ is a $C_k$-coloring of $(G,L')$, contrary to the choice of $c$.
	\end{itemize}
This proves \ref{item:8}.

Let $A=\{v\in V(G)| |L''(v)|=3 \}$. Then for $|L''(v)|\leq 2$ for $v\in G\setminus A$. Pick $v\in A$, then $|L'(v)|=3$ and $v\in Y'$. Let $C$ be the component in $Y'\cup Z'$ contains $v$. By \autoref{clm:update everything}.\ref{item:7} and the construction of $L''$, either  $V(C)\subseteq Y'_i\cup Z'$ for some $i\in [k]$ and $G|(V(C)\cap Y'_i)$ is a stable set, or $V(C)\subseteq Y'_i\cup Y'_{j}$ for some $i,j\in [k]$, $V(C)\cap Y'_i$ and $V(C)\cap Y'_j$ are both non-empty,  and $\{i,j\}\not\subseteq \{\ell,\ell+1\}$ for any $\ell\in [k]$. First assume the latter holds and $v\in Y'_i$, then $v$ is adjacent to $u\in Y'_j$ where $j\notin \{i,i+1,i-1\}$. Recall that $L'$ is reduced on $G|S'\cup X'\cup Y'$. Since  $L'(v)=\{i,i+2,i-2\}$ and $L'(u)\subseteq \{j,j+2,j-2\}$, it follows that $\{i,i+2,i-2\}\subseteq \{j+3,j+1,j-1,j-3\}$. Then since $j\notin\{i-1,i,i+1\}$,   $\{i-2,i,i+2\}=\{j-3,j+1,j+3\}$ or $\{i-2,i,i+2\}=\{j-3,j-1,j+3\}$. In either case, it implies that $k=6$ which is a contradiction. So $V(C)\subseteq Y'_i\cup Z'$ for some $i\in [k]$ and $G|(V(C)\cap Y'_i)$ is a stable set. By \autoref{clm:update everything}.\ref{item:5}, it follows that $N(v)\subseteq X'_i\cup Z$. Let $z\in Z$ be adjacent to $v$. Then by \autoref{clm:Z}, $L(z)$ is a subset of  $ \{i+1,i-1\},\{i+1,i-3\},\{i-1,i+3\}$ or $\{i+3,i-3\}$. This proves \ref{item:9}.
\end{inproof}


Now we can apply \autoref{thm:2SAT} and this completes the proof of correctness of our algorithm. Clearly, the most expensive part of our algorithm is Phase III where
we branch into $O(n^{12k})$ subinstances.  Since each subinstance can be constructed in $O(n^3)$ time by \autoref{lem:reduced list}
and each \sat{2} instance can be solved in $O(n^3)$ time by \autoref{thm:2SAT}, the total running time is  $O(n^{12k+3})$.
\end{proof}
\setcounter{theorem}{7}

\section{Improving the Polynomial Result}\label{sec:polyimp}

Let $k \geq 3$, and let $h : V(G) \rightarrow V(C_k)$ be a homomorphism from a graph $G$ to $C_k$. Let $c_1, \dots, c_k$ denote the vertices of $C_k$ in order. We define a directed graph $G_h$ by assigning a direction to every edge $uv \in E(G)$ as follows: If $h(u) = c_i$ and $h(v) = c_{i+1}$, or if $h(u) = c_k$ and $h(v) = c_1$, we let $uv \in E(G_h)$; otherwise, we let $vu \in E(G_h)$. 

A \emph{walk} in a graph $G$ is a sequence $v_1, \dots, v_j$ of vertices such that for all $i \in \{1, \dots, j-1\}$, $v_iv_{i+1} \in E(D)$.
It is a \emph{closed walk} if in addition, $v_1 = v_j$.
Let $G$ be a graph and let $D$ be an arbitrary orientation of $G$. Note that $D$ has no digons, that is, for all $u, v \in V(D)$, not both $uv \in E(D)$ and $vu \in E(D)$. When talking about walks in $D$, we mean walks in the underlying undirected graph $G$. In particular, walks do not have to preserve directions of edges.
The \emph{slope} $s(W)$ of a walk $W = v_1, \dots, v_j$ in $D$ is defined as $$s(W) = \left|\left\{ i \in \{1, \dots, j-1\} : v_iv_{i+1} \in E(D) \right\}\right| - \left|\left\{ i \in \{1, \dots, j-1\} : v_{i+1}v_i \in E(D) \right\}\right|.$$

\begin{lemma} \label{lem:subpath}
  Let $t, k \in \mathbb{N}$ with $k \geq t+1$. Let $G$ be a connected $P_t$-free graph, and let $h : V(G) \rightarrow V(C_k)$ be a homomorphism. Then $h(V(G))$ is contained in a $(t-1)$-vertex subpath of $V(C_k)$. 
\end{lemma}

\begin{proof}
  Let $G_h$ be as defined above. Let $h'$ be the homomorphism given by the identity map of $C_k$, and let $H = (C_k)_{h'}$. Let $c_1, \dots, c_k$ denote the vertices of $C_k$ in order. We first prove:
  
  \begin{claim} \label{clm:ckwalk}
    Let $W = v_1, \dots, v_j$ be a walk in $H$ with $v_1 = c_a$ and $v_j = c_b$. Then $s(W) + a - b$ is divisible by $k$. 
  \end{claim}

  Suppose for a contradiction that $W$ is a counterexample to Claim \ref{clm:ckwalk} with $j$ minimum. If for some $i \in \{1, \dots, j-2\}$, we have $v_i = v_{i+2}$, then $W' = v_1, \dots, v_i, v_{i+3}, \dots, v_j$ is a walk. Since the walk $W'' = v_i, v_{i+1}, v_{i+2}$ has slope zero, it follows that $s(W') = s(W) - s(W'') = s(W)$, and so $W$ is not a minimum counterexample. It follows that for all $i \in \{1, \dots, j_2\}$, $v_i \neq v_{i+2}$. By symmetry, we may assume that $v_1 = c_a$ and $v_2 = c_{a+1}$. It follows that for all $i \in \{1,\dots, j-1\}$, $v_iv_{i+1} \in E(H)$, and for all $i \in \{1,\dots, j\}$, $v_i = c_{i*}$ where $i-i^*+a-1$ is divisible by $k$. This implies that $s(W) = j-1$, and since $v_j = c_b$, it follows that $j-b+a-1 = s(W) + a - b$ is divisible by $k$, a contradiction. This proves Claim \ref{clm:ckwalk}. 
  
  \begin{claim} \label{clm:div}
    Let $W = v_1, \dots, v_j$ be a walk in $G_h$ with $h(v_1) = h(v_j)$. Then the slope of $W$ is divisible by $k$. 
  \end{claim}

  Let $W' = h(v_1), \dots, h(v_j)$. From the definition of a homomorphism, it follows that $W'$ is a closed walk in $H$. Moreover, from the definition of $H$, it follows that for every edge $uv \in E(G_h)$, we have $h(u)h(v) \in E(H)$; and therefore, $s(W) = s(W')$. Now Claim \ref{clm:div} follows from Claim \ref{clm:ckwalk}.

  \begin{claim} \label{clm:zero}
    Let $W = v_1, \dots, v_j$ be a walk in $G_h$ with $h(v_1) = h(v_j)$. Then the slope of $W$ is $0$. 
  \end{claim}

  Suppose for a contradiction that $W = v_1, \dots, v_j$ is a walk of non-zero slope with $h(v_1) = h(v_j)$, and let $W$ be chosen with $j$ minimum. If there exist $i, i' \in \{1, \dots, j\}$ with $i < i'$, $\{i,i'\} \neq \{1, j\}$ and $v_i = v_{i'}$, then we let $W_1 = v_1, \dots, v_i, v_{i'+1}, \dots, v_j$ and $W_2 = v_i, v_{i+1}, \dots, v_{i'}$. It follows that $W_1$ has the same first and last vertex as $W$, and $W_2$ is a closed walk; and $s(W) = s(W_1) + s(W_2)$. This implies that at least one of $s(W_1), s(W_2)$ is non-zero, contrary to the minimality of $j$. It follows that $v_1, \dots, v_{j-1}$ are distinct, and hence $W$ is the vertex set of a (not necessarily induced) path or cycle $C$ in $G$. 

  Since $s(W) \neq 0$ and $k$ divides $s(W)$, it follows that $C$ has at least $k$ vertices. Since $k \geq t+1$, and $G$ is $P_t$-free, it follows that $C$ is not an induced path and not an induced cycle. Let $uw \in E(G)$ such that $u, w \in V(C)$, but $\{u, w\} \neq \{v_1, v_j\}$ and there is no $i \in \{1, \dots, j-1\}$ such that $\{u, w\} = \{v_i, v_{i+1}\}$. By symmetry, we may assume that $u = v_i, w = v_{i'}$ and $i < i'$. Now let $W_3 = v_1, \dots, v_i, v_{i'}, \dots, v_j$ and $W_4 = v_i, v_{i+1}, \dots, v_{i'}, v_i$. It follows that $s(W) = s(W_3) + s(W_4)$, since each consecutive pair of vertices of $W$ occurs in exactly one of $W_3, W_4$; and the pair $v_i, v_{i'}$ occurs in opposite orders in $W_3$ and $W_4$. This implies that at least one of $s(W_3), s(W_4)$ is non-zero. From the choice of $u$ and $w$, it follows that both $W_3$ and $W_4$ have fewer vertices than $W$, a contradiction. This implies Claim \ref{clm:zero}.

  \medskip

  Let us say that $u$ \emph{precedes} $w$ if $u, w \in V(G_h)$ and there is a walk $W = v_1, \dots, v_j$ with $v_1 = u$ and $v_j = w$ such that $s(W) > 0$ in $G_h$. From Claim \ref{clm:zero}, it follows that no vertex precedes itself. Moreover, the definition immediately implies that this property is transitive, that is, if $u$ precedes $w$ and $w$ precedes $y$, then $u$ precedes $y$. This defines a strict partial order; and hence there is a vertex $u \in V(G_h)$ such that no vertex precedes $u$. By symmetry, we may assume that $h(u) = c_1$. Now suppose that there is a vertex $w \in V(G_h)$ with $h(w) = c_l$ for some $l \in \{t, \dots, k\}$. Since $G$ is connected, it follows that there is an induced $u$-$w$-path $P$ in $G$, for example by taking $P$ to be a shortest $u$-$w$-path. Let $W = v_1, \dots, v_j$ denote the vertices of $P$ in reverse order; it follows that $W$ is a walk with $h(v_1) = c_l$ and $h(v_j) = c_1$. From Claim \ref{clm:ckwalk}, we deduce that $k$ divides $s(W) + l - 1$, and since $0 < |l-1| < k$, it follows that $s(W) \neq 0$. Since $P$ has length at most $t-2$, it follows that $|s(W)| \leq t-2 \leq k-3$. This implies that $s(W) +l -1 \in \{0, k\}$. Since $l - 1 \geq t-1 > |s(W)|$, it follows that $s(W) = k - l + 1 > 0$. This implies that $w$ precedes $u$, contradicting the choice of $u$. It follows that $h(V(G)) \subseteq \{c_1, \dots, c_{t-1}\}$, as claimed.
\end{proof}

This implies the following:

\begin{theorem} 
  Let $G$ be connected $P_t$-free graph. Then, for all $k, k' > t$, $G$ has a $C_k$-coloring if and only if $G$ has a $C_{k'}$-coloring. 
\end{theorem}

It also leads to the following improvement of Theorem \ref{thm:polynomial}.
\thmimprove*
\begin{proof}
If $k < 10$, then this follows from Theorem \ref{thm:polynomial}.
If $k \geq 10$, then Lemma \ref{lem:subpath} implies that every $C_k$-coloring of $G$ is contained in an 8-vertex subpath of $C_k$.
Since there are $k$ such subpaths and \lcol{P_t} is polynomial-time solvable for every $t$ even in general graphs~\cite{DBLP:journals/combinatorica/FederHH99},  the result follows. 
\end{proof}

\section{Hardness results}\label{sec:NPc}
In this section we study the complexity of variants of \col{C_k} in $F$-free graphs, if $F$ is not a path.
First we consider the case of odd $k$, and then the case of even $k$.

\subsection{Complexity of variants of \col{C_k} for odd $k \geq 5$}
Recall that \col{C_k} is NP-complete for every odd $k \geq 3$~\cite{HN90}. In this section we prove the following theorem.

\thmhard*

We will prove \autoref{thm:main-hardness} in several steps in which we analyze possible structure of $F$.
We start with the following simple observation that will be repeatedly used. For the rest of this section, let $k=2s+1$ for $s\geq 2$.

\begin{observation}\label{ob:simple}
Let $s \geq 2$ be an integer and $P$ be a $2s$-vertex path with endvertices $a$ and $b$.
Then the following holds.
\begin{itemize}
\item In any $C_{2s+1}$-coloring $h$ of $P$ we have $h(a) \neq h(b)$.
\item For any distinct $i,j \in \{1,2,\ldots,2s+1\}$, there exists a $C_{2s+1}$-coloring $h$ of $P$ such that $h(a) = i$ and $h(b)=j$. \qedhere
\end{itemize}
\end{observation}

\subsubsection*{Eliminate cycles}

The \emph{girth} of a graph $G$, denoted by $\girth(G)$, is the length of a shortest cycle in $G$.
A vertex in a graph is called a {\em branch vertex} if its degree is at least 3.
By $\Gamma_p$ we denote the class of graphs, in which the number of edges in any path joining two branch vertices is divisible by $p$.

We first show that the problem is NP-hard in $F$-free graphs, unless $F$ is a tree in $\Gamma_{2s-1}$.

\begin{theorem}\label{thm:cycle}
For each fixed integer $s\ge 2$ and each connected graph $F$, \col{C_{2s+1}}  is NP-complete for $F$-free graphs 
whenever $F$ contains a cycle or is not in $\Gamma_{2s-1}$.
\end{theorem}

\begin{proof}
It is known (see e.g. \cite{KL07}) that 
the \col{(2s+1)} problem is NP-complete for graphs of girth at least $g$ for each fixed $g\ge 3$.
We reduce this problem to \col{C_{2s+1}}. Given a graph $G$, we obtain a graph $G'$ by replacing each edge of $G$ by a $(2s-1)$-edge path.
Then it follows from \autoref{ob:simple} that $G$ is $(2s+1)$-colorable if and only if $G'$ is $C_{2s+1}$-colorable. 
Indeed, by the first bullet point in \autoref{ob:simple} we observe that any $C_{2s+1}$-coloring of $G'$, restricted to the vertices of $G$,
is a proper $(2s+1)$-coloring. On the other hand, by the second bullet point, any proper $(2s+1)$-coloring of $G$ can be extended to a $C_{2s+1}$-coloring of $G'$.

Clearly, $\girth(G') = \girth(G) \cdot (2s-1) \geq g(2s-1)$.
Thus, if we choose $g\ge 3$ such that $g(2s-1)>\girth(F)$, e.g., $g = |V(F)|+1$, it follows that all graphs of girth at least $g(2s-1)$ are $F$-free. Moreover, it is easy to see that the number of edges in any path joining two branch vertices of $G'$ is divisible by $2s-1$, so if $F \notin \Gamma_{2s-1}$, then $G'$ does not contain $F$. \end{proof}

\subsubsection*{Eliminate vertices of degree at least 4}

From now on it suffices to consider trees with branch vertices at distance divisible by $2s-1$.
We now show that \col{C_k} is NP-complete for $F$-free graphs if $F$ contains a vertex of degree at least 4.
Note that in this case every subcubic graph is $F$-free.

\begin{theorem}\label{thm:maximum degree}
For each fixed $s\ge 2$, \col{C_{2s+1}} is NP-complete for subcubic graphs.
\end{theorem}

\begin{proof}
We reduce from \col{C_{2s+1}} for general graphs. 
Let $G$ be a graph. We construct an equivalent instance of \col{C_{2s+1}} with maximum degree 3 as follows. 
If $\Delta(G) \leq 3$, we are done. Otherwise, let $v$ be a vertex of degree $d \geq 4$, and let $u_1,\ldots,u_d$ be its neighbors.
we replace $v$ with a copy of $R^d$ (see \autoref{fig:VR}).
More specifically, we start the construction of $R^d$ by introducing $d$ copies of $C_{2s+1}$;
let the vertices of the $i$-th copy be denoted by $v_{1,i},v_{2,i},\ldots,v_{2s-1,i}$.
Then, for each $i < d$, we identify the vertex $v_{2,i}$ with the vertex $v_{2s-1,i+1}$ and the vertex $v_{3,i}$ with the vertex $v_{2s-2,i-1}$.
The vertices $v_{1,1},v_{1,2},\ldots,v_{1,d}$ are called \emph{output vertices}, each of them becomes adjacent to a distinct vertex from $u_1,\ldots,u_d$.

Since any homomorphism from $C_{2s+1}$ to itself must be an isomorphism, it follows from the definition of homomorphism that in any $C_{2s+1}$-coloring of $R^d$, each output vertex must be mapped to the same vertex of  $C_{2s+1}$.
Therefore, the obtained graph is $C_{2s+1}$-colorable if and only if the original one is. 
Moreover, the number of vertices of degree greater than 3 in the new graph is one less than that in $G$.
Therefore, by repeating this procedure exhaustively, we finally obtain a subcubic graph, which is an equivalent instance of \col{C_{2s+1}}.
\end{proof}

\begin{figure}[h!]
\centering
\begin{tikzpicture}[scale = 0.5]
\tikzstyle{vertex}=[draw, circle, fill=black!10,inner sep=1.5pt]

\foreach \i in {0,2,4,8,10,12}
{
	\node[vertex, fill = black] (1\i) at (\i,0) {};
	\node[vertex] (2\i) at (\i+1,-1) {};
	\node[vertex] (3\i) at (\i+1,-2) {};
	\node[vertex] (4\i) at (\i+0.5,-3) {};
	\node[vertex] (5\i) at (\i-0.5,-3) {};
	\node[vertex] (6\i) at (\i-1,-2) {};
	\node[vertex] (7\i) at (\i-1,-1) {};

	\draw (1\i) -- (2\i) -- (3\i);
	\draw (4\i) -- (5\i);
	\draw (6\i) -- (7\i) -- (1\i);
	
	\draw[densely dashed] (3\i) -- (4\i);
	\draw[densely dashed] (5\i) -- (6\i);	
}

\draw[dotted] (5.5,-1) --++ (1,0);
\draw[dotted] (5.5,-2) --++ (1,0);
\draw[dotted] (5.5,-3) --++ (1,0);

\draw[decoration={brace,mirror,raise=5pt},decorate]
  (-1,-3.1) -- node[below=6pt] {$d$ copies of $C_{2s+1}$} (13,-3.1);
  
%
  
\end{tikzpicture}
\caption{The graph $R^d$ that consists of a chain of $d$ copies of $C_{2s+1}$. The vertices marked black are called {\em output vertices}.}\label{fig:VR}
\end{figure}

\subsubsection*{Eliminate multiple branch vertices}

Before we prove the main theorem we need one more intermediate step that allows us to eliminate those $F$ in which 
there are two branch vertices that are at distance not divisible by $s$.
The proof is a reduction from the problem called {\sc Non-Rainbow Coloring Extension}, whose instance is a 3-uniform hypergraph $H$ and a partial coloring $f$ of some of its vertices with colors $\{1,2,3\}$. We ask whether $f$ can be extended to a 3-coloring of $V(H)$ such that no hyperedge is {\em rainbow} (i.e., contains three distinct colors). This problem is known to be NP-complete~\cite{BODIRSKY20121680}.


\begin{theorem}\label{thm:bipartiteishard}
For each fixed integer $s\ge 2$, \ext{C_{2s+1}} is NP-complete for bipartite graphs in $\Gamma_s$.
\end{theorem}

\begin{proof}
We reduce from  {\sc Non-Rainbow Coloring Extension}.
Let $H=(V,E)$ be a 3-uniform hypergraph and let $f$ be a partial 3-coloring of $H$.
We construct an instance of \ext{C_{2s+1}} as follows.
\begin{itemize}
\item For each vertex $v \in V$, we introduce a variable vertex, denoted by $v'$. 
If $v$ is precolored by $f$, we precolor $v'$ with the color $f(v)$.
\item For each $v$ that is not precolored by $f$, we introduce $2s-2$ new vertices and 
precolor them with $4,5,\ldots,2s+1$, respectively.  Then each of these new vertices is joined by a $(2s-1)$-edge path to $v'$.
It follows from \autoref{ob:simple} that each vertex $v'$ can only be mapped to one of $1,2,3$, and any of these three choices is possible.
\item For each hyperedge $e=\{x,y,z\}\in E$, we add a new vertex $v_e$ and 
three $s$-edge paths connecting $v_e$ to $x'$,$y'$, and $z'$, respectively. 
This whole subgraph is called an \emph{edge gadget}. 
\end{itemize}
Observe that if $x'$ is mapped to $i \in \{1,2,3\}$, then the possible colors for $v_e$ 
are $\{s+i, s+i-2,\ldots, s+i-2\lfloor s/2 \rfloor\} \cup \{s+i+1, s+i+3,\ldots, s+i+1+2\lfloor s/2 \rfloor\}$. 
Thus, if each of $x',y',z'$ is mapped to a different vertex from $\{1,2,3\}$, then there is no way to extend this mapping to the whole edge gadget. 
On the other hand, such an extension is possible whenever $x',y',z'$ receive at most two distinct colors.

We denote by $G$ the resulting graph. By the properties of variable vertices and edge gadgets, 
$(H,f)$ is an yes-instance of {\sc Non-Rainbow Coloring Extension} if and only if the precoloring of $G$ can be extended to a $C_{2s+1}$-coloring of $G$.
Clearly, $G$ is bipartite and belongs to $\Gamma_s$.
\end{proof}

Let us summarize the cases that are covered by Theorems \ref{thm:cycle}, \ref{thm:maximum degree}, and \ref{thm:bipartiteishard}.
Consider the \ext{C_{2s+1}} in $F$-free graphs, where $F$ is a fixed connected graph.
If $F$ has a cycle, then the problem is NP-hard by Theorem \ref{thm:cycle}.
If $F$ has a vertex of degree at least 4, then the problem is NP-hard by Theorem \ref{thm:maximum degree}.
Thus let $F$ be a subcubic tree.
If $F$ has two branch vertices whose distance is not divisible by $s$, then the problem is NP-hard by Theorem \ref{thm:bipartiteishard}.
If $F$ has two branch vertices at distance not divisible by $2s-1$, then the problem is NP-hard by Theorem \ref{thm:cycle}.
Summing up, the only cases left are subcubic trees where all pairs of branch vertices are at pairwise distance divisible by $s$ and by $2s-1$.
In other words, we are left with the case that $F$ is a subcibic tree in $\Gamma_{s(2s-1)}$ (observe that $s$ and $2s-1$ are relatively prime).

In the next step we show that the problem is NP-hard if $F$ has more than one branch vertex.

\begin{theorem} \label{thm:onebranch}
Let $s \geq 2$ be an integer and let $F$ be a tree in $\Gamma_{s(2s-1)}$.
If $F$ contains two branch vertices, then \col{C_{2s+1}} is NP-complete for $F$-free graphs.
\end{theorem}

\begin{proof}
Let $d$ be the distance between two closest branch vertices in $F$.
We reduce from {\sc Positive Not-All-Equal Sat} with all clauses containing exactly three literals. 
Consider an instance with variables $x_1,x_2,\ldots,x_n$ and clauses $D_1,D_2,\ldots,D_m$.
\begin{itemize}
\item We start our construction by introducing one special vertex $z$. 
\item For each variable $x_i$, we introduce a vertex $v_i$, adjacent to $z$.
\item For each clause $D_\ell = \{x_i,x_j,x_k\}$, we introduce three new vertices $y_{\ell,i}$, $y_{\ell,j}$, and $y_{\ell,k}$, 
and join each pair of them with a $(2s-1)$-edge path. This guarantees that in every $C_{2s+1}$-coloring, 
they get three distinct colors. These three paths constitute the \emph{clause gadget}.
\item For each variable $x_i$ belonging to a clause $D_\ell$, we join each $y_{\ell,i}$ to $v_i$ by a path $P_{\ell,i}$ 
with $2d(2s-1)+1$ edges. Let $v_i = p_1,p_2,\ldots,p_{2d(2s-1)+2} = y_{\ell,i}$ be the consecutive vertices of $P_{\ell,i}$.
We add edges joining $z$ and $p_{1+j(2s-1)}$ for every $1 \leq j \leq 2d$.
\end{itemize}
This completes the construction of a graph $G$. We claim that $G$ is $C_{2s+1}$-colorable if and only if the initial formula is satisfiable, and that $G$ belongs to our class.

\begin{claim}\label{clm:onebranch-colorable}
$G$ is $C_{2s+1}$-colorable if and only if the initial formula is satisfiable.
\end{claim}

\begin{inproof}
Suppose first that the formula has a satisfying assignment $\sigma$. 
We color the vertex $z$ with color $2$. If a variable $x_i$ is set true by $\sigma$, we color $v_i$ with color 1,  
otherwise we color $v_i$ with color 3.
Let us consider an arbitrary clause $D_\ell = \{x_i,x_j,x_k\}$. 
We extend this coloring to all paths $P_{\ell,i}$ ($P_{\ell,j}$, $P_{\ell,k}$, accordingly), 
so that the color of $p_{1+j(2s-1)}$ (for every even $2 \leq j \leq 2d$) is the same as the color of $p_1$.
Therefore, if $v_i$ is colored 1, then the possible colors for $y_{\ell,i}$ are 2 and $2s+1$, 
and if $v_i$ is colored 3, then the possible colors for $y_{\ell,i}$ are 2 and 4. 
Since $D_\ell$ contains at least one true variable and at least one false variable, 
we can choose three distinct colors for $y_{\ell,i}, y_{\ell,j}$, and $y_{\ell,k}$, and extend this mapping to the remaining vertices.

Suppose now that there exists a $C_{2s+1}$-coloring $f$ of $G$. 
By symmetry, we may assume that $f(z)=2$.  This implies that every $v_i$ is colored by 1 or 3. 
We define the assignment: $\sigma(x_i)$ is true if $f(v_i)=1$ and false otherwise. 
Suppose that $\sigma$ is not satisfying, i.e., there is a clause $D_\ell$ with literals $x_i, x_j, x_k$ that all have the same value. 
It follows that $f(v_i) = f(v_j) = f(v_k)$ and without loss of generality, we may assume that $f(v_i) = 1$.  
Observe that for every even $2 \leq j \leq 2d$ we have $f(p_{1+j(2s-1)})=f(v_i) = 1$, 
where $p_t$'s are consecutive vertices of $P_{\ell,i}$. This implies that $f(y_{\ell, i}) \in \{2, 2s+1\}$. 
Similarly, $f(y_{\ell, j}), f(y_{\ell, k}) \in \{2, 2s+1\}$. 
It follows that there are two of $i, j, k$, say $i$ and $j$, such that $f(y_{\ell,i}) = f(y_{\ell,j})$. 
But these two vertices are connected by a path with $2s-1$ edges, which contradicts \autoref{ob:simple}. 
\end{inproof}

Moreover, the constructed graph belongs to our class.
\begin{claim}\label{clm:onebranch-ffree}
$G$ is $F$-free.
\end{claim}

\begin{inproof}
Let $a$ and $b$ two branch vertices in $F$, they are at distance $d$.
Assume by contradiction that $G$ contains an induced copy of $F$.
Let  $g : V(F) \rightarrow V(G)$ map every vertex of $F$ to its corresponding vertex in an induced copy of $F$ in $G$.

Suppose first that $z \in g(V(F))$. Since $d$ is divisible by $s(2s-1)$, it follows that $a$ and $b$ have distance at least 6 in $F$. 
Therefore, for every vertex $u \in V(F)$, there is a branch vertex in $F \setminus (N(u) \cup \{u\})$. 
Now let $u \in V(F)$ such that $g(u) = z$. Then $F \setminus (N(u) \cup \{u\})$ is an induced subgraph of $G \setminus (N(z) \cup \{z\})$, 
but the latter graph has maximum degree two, a contradiction. 
It follows that $z \not\in g(V(F))$,  and so $F$ is an induced subgraph of $G' = G \setminus \{z\}$. 

Let us now consider the possible values of $g(a)$ and $g(b)$. 
Since $a$ and $b$ have degree at least 3 in $F$, it follows that $g(a)$ and $g(b)$ have degree at least 3 in $G'$. 
Moreover, since $a$ and $b$ are at distance $d$ in $F$, it follows that $g(a)$ and $g(b)$ are at distance at most $d$ in $G'$.

\medskip
\noindent {\bf Case 1.} $g(a) = v_i$ or $g(b) = v_i$ for some $i$.
Every vertex $u \neq v_i$ of degree at least three in $G'$ has distance at least $2d(2s-1)+1 > d$ from $v_i$, a contradiction.

\medskip
\noindent {\bf Case 2.} $g(a) = y_{\ell,i}$ for some $i$ and $\ell$.
By the first case, it follows that $g(b) = y_{\ell', j}$ for some $j$ and $\ell'$. 
Let $Q$ be the $a$-$b$-path in $F$. Since $Q$ has $d$ edges, and since the number of edges of every path in $G'$ 
between $y_{\ell, i}$ and $y_{\ell', j}$ for $\ell\neq \ell'$ is more than $d$, it follows that $\ell = \ell'$ and $g(V(Q))$ is contained in the clause gadget for $D_{\ell}$. 

Suppose first that $s \geq 3$. Since $d$ is divisible by $s(2s-1)$, it follows that $Q$ has $d+1 \geq 3(2s-1)+1$ vertices.
However,  the clause gadget for $D_{\ell}$ has $3(2s-1)$ vertices, a contradiction. 

Therefore, $s = 2$. Then the clause gadget for $D_{\ell}$ is isomorphic to a nine-cycle in $G'$ with vertices $c_1, \dots, c_9$ in this order, say. 
By symmetry, we may assume that $g(a) = c_1$ and $g(b) = c_4$. 
Since $d$ is divisible by $s(2s-1)$ and $s = 2$ , it follows that $d = 6$, and so $g(V(Q)) = \{c_1, c_4, c_5, c_6,c_7,c_8,c_9\}$.  
Since $F$ is a tree, it follows that either $c_2 \not\in g(V(F))$ or $c_3 \not\in g(V(F))$. By symmetry, we may assume that $c_2 \not\in g(V(F))$. 
But $c_1$ has degree two in $G' \setminus \{c_2\}$, a contradiction. This concludes the proof of the claim.
\end{inproof}

This completes the proof of \autoref{thm:onebranch}.
\end{proof}

Now \autoref{thm:main-hardness} comes from combining the Theorems \ref{thm:cycle}, \ref{thm:maximum degree}, \ref{thm:bipartiteishard}, and \ref{thm:onebranch}.
\subsection{Complexity of variants of \col{C_k} for even $k$}
Recall that that in case of even $k$, the \col{C_k} problem is polynomial-time solvable for general graphs~\cite{HN90}.
However, this is no longer true in the case of \lcol{C_k}: the problem is polynomial-time solvable for $k=4$ and NP-complete for all even $k \geq 6$~\cite{DBLP:journals/combinatorica/FederHH99}.
For the rest of this section $k=2s$, where $s \geq 3$. The consecutive vertices of $C_{2s}$ are denoted by $\{1,2,\ldots,2s\}$ (with $2s$ adjacent to $1$).

In this section we prove \autoref{thm:main-hardness-even}.

\thmevenhard*

To be more specific, we will prove the following.
\begin{theorem}\label{thm:even-hard}
Let $g$ be a fixed integer. For every even $k \geq 6$ the \lcol{C_k} problem is NP-complete for subcubic graphs with girth at least $g$, in which every pair of branch vertices is at distance at least $g$.
\end{theorem}
\begin{proof}
We will reduce from \msat{3}, a variant of \sat{3}, in which every clause is of size at most 3 and contains only positive or only negative literals.
It is known that this problem is NP-complete, even if each variable appears at most three times~\cite{doi:10.1142/S0129054118500168}.

For every variable $v_i$ we introduce a \emph{variable vertex} $x_i$ with list $\{1,3\}$. Color 1 will correspond to true assignment, while 3 will denote false. For every clause $D_\ell$, we introduce a \emph{clause vertex} $d_\ell$ with list $\{1,3,5\}$.

Now we need to connect variable vertices with clause vertices. 
For $i = \{1,3,5\}$ we will construct a path $Q^{(i)}$, starting in a vertex $a^i$ and ending in a vertex $b^i$, with appropriately chosen lists, so that the following are satisfied:
\begin{enumerate}
\item for $i \in \{1,3,5\}$, if $a^i$ is colored 1, then for every $c \in \{1,3,5\}$ there is a list $C_{2s}$-coloring of $Q^{(i)}$ in which the color of $b^i$ is $c$,
\item for $i \in \{1,3,5\}$ and $c \in \{1,3,5\} \setminus \{i\}$ there is a list $C_{2s}$-coloring of $Q^{(i)}$ in which the color of $a^i$ is 3 and the color of $b^i$ is $c$,
\item for $i \in \{1,3,5\}$ there is no list $C_{2s}$-coloring of $Q^{(i)}$ in which $a^i$ is colored 3 and $b^i$ is colored $i$.
\end{enumerate}
Additionally, we will make sure that each path has more than $g$ vertices.

Suppose for now that we have constructed such paths. Consider a clause with three positive literals, $D_\ell = (v_p,v_q,v_r)$ and let the ordering of variables be fixed. We introduce a copy of each $Q^{(i)}$ for $i \in \{1,3,5\}$.
We identify the vertex $a^1$ with $x_p$, the vertex $a^3$ with $x_q$, and the vertex $a^5$ with $x_r$. Finally, we identify the vertices $b^1,b^3,b^5$, and $d_\ell$.
In case of a clause $D_\ell$ with two positive literals, we use only paths $Q^{(1)}$ and $Q^{(3)}$, and remove the color $5$ from the list of $d_\ell$.

It is clear that the paths ensure that in order to find a color for $d_\ell$, at least one of corresponding variable vertices must be colored 1, meaning that one of the variables in $D_\ell$ is true.

In order to deal with clauses with negative literals, for $i = \{1,3,5\}$ we will construct a path $\overline{Q}^{(i)}$, starting in a vertex $\overline{a}^i$ and ending in a vertex $\overline{b}^i$, with appropriately chosen lists, so that the following are satisfied:
\begin{enumerate}
\item for $i \in \{1,3,5\}$, if $\overline{a}^i$ is colored 3, then for every $c \in \{1,3,5\}$ there is a list $C_{2s}$-coloring of $\overline{Q}^{(i)}$ in which the color of $\overline{b}^i$ is $c$,
\item for $i \in \{1,3,5\}$ and $c \in \{1,3,5\} \setminus \{i\}$ there is a list $C_{2s}$-coloring of $\overline{Q}^{(i)}$ in which the color of $\overline{a}^i$ is 1 and the color of $\overline{b}^i$ is $c$,
\item for $i \in \{1,3,5\}$  there is no list $C_{2s}$-coloring of $\overline{Q}^{(i)}$ in which $\overline{a}^i$ is colored 1 and $\overline{b}^i$ is colored $i$.
\end{enumerate}

Now, for a clause $D_\ell = (\lnot v_p, \lnot v_q, \lnot v_r)$, we introduce a copy of each $\overline{Q}^{(i)}$ for $i \in \{1,3,5\}$, and identify the vertex $\overline{a}^1$ with $x_p$, the vertex $\overline{a}^3$ with $x_q$, the vertex $\overline{a}^5$ with $x_r$, and finally all vertices $\overline{b}^1,\overline{b}^3,\overline{b}^5$, and $d_\ell$.
Similarly, for a clause $D_\ell = (\lnot x_p, \lnot x_q)$ we use paths $\overline{Q}^{(1)}$ and $\overline{Q}^{(3)}$ and remove the color 5 from the list of $d_\ell$.

It is straightforward to observe that the constructed graph $G$ admits a list $C_{2s}$-coloring if and only if the initial \msat{3} formula is satisfiable.

Now let us argue that $G$ belongs to the considered class. Note that the only branch vertices in $G$ are variable vertices and clause vertices. Since each clause contains at most three variables and each variable appears in at most 3 clauses, we conclude that $G$ is subcubic. Finally, since all paths joining variable vertices with clause vertices have more than $g$ vertices, we conclude that $G$ has no cycle of length at most $g$.

Thus in order to complete the proof, we need to show how to construct $Q^{(i)}$'s and $\overline{Q}^{(i)}$'s.
Let us assume that $g$ is even (we can do it safely, as we can always replace $g$ with $g+1$ and the claim will still hold). The lists of consecutive vertices on $Q^{(1)}$ are as follows (starting from $a^1$):
\begin{align*}
\underbrace{\{1,3\}, \{2s,4\}, \{1,3\}, \{2s,4\}, \ldots, \{1,3\}, \{2s,4\}}_{g},\{1,3\},\{2s,2\},\{2s-1,3\},\{2s,4\},\{1,3,5\}.
\end{align*}
The lists of consecutive vertices on $Q^{(5)}$ are as follows  (starting from $a^5$):
\begin{align*}
\underbrace{\{1,3\}, \{2s,4\}, \{1,3\}, \{2s,4\}, \ldots, \{1,3\}, \{2s,4\}}_{g},\{1,3\},\{2s,2\},\{2s-1,3\},\ldots,\{6,2\},\{1,3,5\}.
\end{align*}
The construction of $Q^{(3)}$ is slightly more complicated.
For $s=3$, the lists of consecutive vertices are as follows (starting from $a^3$):
\begin{align*}
\underbrace{\{1,3\}, \{4,6\}, \{1,3\},  \ldots, \{4,6\}}_{g}, \{1,3\}, \{2,6\}, \{1,5\}, \{4,6\}, \{1,3,5\}.
\end{align*}
For $s \geq 4$, the lists of consecutive vertices  are as follows:
\begin{align*}
& \underbrace{\{1,3\}, \{2s,4\}, \{1,3\}, \{2s,4\}, \ldots, \{1,3\}, \{2s,4\}}_{g},\{1,3\},\{2s,2\},\{2s-1,3\},\{2s,4\},\\
& \{1,5\}, \{2,6\}, \{3,5,7\}, \{2,6,8\}, \{3,5,9\}, \ldots, \{2,6,2s\},\{1,3,5\}.
\end{align*}

It is straightforward to verify that they satisfied the required conditions. 
The construction of the paths $\overline{Q}^{(i)}$ for $i = \{1,3,5\}$ is analogous, with the roles of $1$ and $3$ switched. This completes the proof.
\end{proof}

Now let $F$ be a connected graph that is not a subdivided claw. \autoref{thm:main-hardness-even} follows by applying \autoref{thm:even-hard} with $g = |F|+1$.

\subsection{Subexponential algorithms and ETH-based lower bounds}

Recall that by the result of Groenland {\em et al.}~\cite{GORSSS18}, for every fixed $t$ and $k$, the \lcol{C_k} problem can be solved in time $2^{O(\sqrt{n \log n})}$ for $P_t$-free graphs. 

It turns out that such subexponential algorithms for variants of \col{C_k} are unlikely to exist for $F$-free graphs, if $F$ is not a subgraph of a subdivided claw.
All reductions in our hardness proofs are linear in the number of vertices (recall that the target graph is assumed to be fixed). Moreover, all problems we are reducing from can be shown to be NP-complete by a linear reduction from \sat{3}. Thus we get the following results, conditioned on the Exponential Time Hypothesis (ETH), which, along with the \emph{sparsification lemma}, implies that \sat{3} with $n$ variables and $m$ clauses cannot be solved in time $2^{o(n+m)}$~\cite{ImpagliazzoPaturi,DBLP:journals/jcss/ImpagliazzoPZ01}.

\begin{corollary}
Unless the ETH fails, the following holds. If $F$ is a connected graph 
that is not a subgraph of a subdivided claw, then for every $s \geq 2$, the problems
\begin{enumerate}[a)]
\item  \ext{C_{2s+1}} and
\item  \lcol{C_{2s+2}}
\end{enumerate}
cannot be solved in time $2^{o(n)}$ in $F$-free graphs with $n$ vertices.
\end{corollary}

\section{Conclusion}\label{sec:conclusion}
In this paper, we initiate a study of variants \col{C_{k}} for $F$-free graphs for a fixed graph $F$.
We show that \lcol{C_{k}} is polynomial-time solvable for $P_9$-free graphs, whenever $k = 5$ or $k = 7$ or $k \geq 9$.
Moreover, we prove that for every $s \geq 2$ the \ext{C_{2s+1}} and \lcol{C_{2s+2}} problems are NP-complete for $F$-free graphs if some component of $F$ is not a subdivided claw.
Note that for the case if $k$ is odd, all our hardness results work for \col{C_{2s+1}}, except for \autoref{thm:bipartiteishard}.
Thus it is natural to ask whether an analogous hardness result holds for \col{C_{2s+1}} too.

Moreover, the following questions seem natural to explore.
\begin{itemize}
\item Are there values of $s$ and $t$ such that \col{C_{2s+1}} is NP-complete for $P_t$-free graphs?
\item Are there values of $k$ and $t$ such that \lcol{C_{k}} is NP-complete for $P_t$-free graphs?
\item Is \col{C_{2s+1}} polynomial for $F$-free graphs when $F$ is a subdivided claw?
\end{itemize}

We also believe that it would be interesting to study the complexity of \ext{C_k} for even $k$ on restricted classes of graphs. 
It is known that this problem is NP-complete (in general graphs) for every $k \geq 6$~\cite{DBLP:journals/combinatorica/FederHH99}.
Recall that  \lcol{C_4} (and thus \ext{C_4}) is polynomial-time solvable in general graphs~\cite{DBLP:journals/combinatorica/FederHH99}.

Let us point out that, quite surprisingly, \ext{C_6} is polynomial-time solvable for graphs with maximum degree 3~\cite{DBLP:journals/dam/FederHH09}. Since every graph that contains a triangle is clearly a no-instance of \ext{C_6}, we conclude that the problem is polynomial-time solvable for $K_{1,4}$-free graphs.

On the other hand, it is known that \ext{C_6} is NP-complete for graphs with maximum degree 4, and for every even $k \geq 8$, \ext{C_k} is NP-complete for graphs with maximum degree 3~\cite{DBLP:journals/dam/FederHH09}.

Finally, note that the list of each vertex $v$ in the construction in the proof of \autoref{thm:main-hardness-even} can be simulated by introducing some number of precolored vertices and joining them to $v$ with paths of certain length. Since the newly introduced vertices are private for every original vertex, we do not introduce any new cycles, thus the constructed graph still has large girth. Thus we obtain the following.

\begin{corollary}
Let $F$ be a connected graph and $k \geq 8$ be an even integer. The following problems are NP-complete for $F$-free graphs:
\begin{enumerate}[a)]
\item the \ext{C_6} problem, if $F$ is not a tree with $\Delta(F) \leq 4$,
\item the \ext{C_k} problem, if $F$ is not a tree with $\Delta(F) \leq 3$.
\end{enumerate} 
\end{corollary}

Let us also point out that the reduction described above introduces many further constraints on $F$. Moreover, if we modify the construction so that the precolored vertices used for simulating lists are not private, but shared by original vertices, further constrains can be introduced. However, the description of forbidden subgraphs is not elegant and we believe that it can be further improved.

\paragraph{Follow-up work.} During the reviewing process of this paper, Okrasa and Rz\k{a}\.zewski~\cite{DBLP:conf/stacs/OkrasaR21} considered the complexity of \lcol{H} in $F$-free graphs. Among other results, they showed that for every $k \geq 5$, the \lcol{C_k} problem in $F$-free graphs can be solved in \emph{quasipolynomial} time if $F$ is a path, and in subexponential time if $F$ is any subdivision of the claw.

\bibliography{main}

\begin{thebibliography}{10}

\bibitem{alekseev1982effect}
Vladimir~E Alekseev.
\newblock The effect of local constraints on the complexity of determination of
  the graph independence number.
\newblock {\em Combinatorial-algebraic methods in applied mathematics}, pages
  3--13, 1982.

\bibitem{APT79}
Bengt Aspvall, Michael~F. Plass, and Robert~Endre Tarjan.
\newblock A linear-time algorithm for testing the truth of certain quantified
  boolean formulas.
\newblock {\em Inf. Process. Lett.}, 8(3):121--123, 1979.

\bibitem{DBLP:journals/algorithmica/BacsoLMPTL19}
G{\'{a}}bor Bacs{\'{o}}, Daniel Lokshtanov, D{\'{a}}niel Marx, Marcin
  Pilipczuk, Zsolt Tuza, and Erik~Jan van Leeuwen.
\newblock Subexponential-time algorithms for maximum independent set in
  ${P}_t$-free and broom-free graphs.
\newblock {\em Algorithmica}, 81(2):421--438, 2019.

\bibitem{BODIRSKY20121680}
Manuel Bodirsky, Jan Kára, and Barnaby Martin.
\newblock The complexity of surjective homomorphism problems -- a survey.
\newblock {\em Discrete Applied Mathematics}, 160(12):1680 -- 1690, 2012.

\bibitem{BCMSSZ17}
Flavia Bonomo, Maria Chudnovsky, Peter Maceli, Oliver Schaudt, Maya Stein, and
  Mingxian Zhong.
\newblock Three-coloring and list three-coloring of graphs without induced
  paths on seven vertices.
\newblock {\em Combinatorica}, 38(4):779--801, 2018.

\bibitem{DBLP:conf/focs/Bulatov17}
Andrei~A. Bulatov.
\newblock A dichotomy theorem for nonuniform {CSPs}.
\newblock In Umans \cite{DBLP:conf/focs/2017}, pages 319--330.

\bibitem{CS16}
Eglantine Camby and Oliver Schaudt.
\newblock A new characterization of {$P_k$}-free graphs.
\newblock {\em Algorithmica}, 75(1):205--217, 2016.

\bibitem{DBLP:conf/soda/2019}
Timothy~M. Chan, editor.
\newblock {\em Proceedings of the Thirtieth Annual {ACM-SIAM} Symposium on
  Discrete Algorithms, {SODA} 2019, San Diego, California, USA, January 6-9,
  2019}. {SIAM}, 2019.

\bibitem{DBLP:conf/esa/ChudnovskyHRSZ19}
Maria Chudnovsky, Shenwei Huang, Pawe{\l} Rz{\k{a}}{\.{z}}ewski, Sophie Spirkl,
  and Mingxian Zhong.
\newblock Complexity of c\({}_{\mbox{k}}\)-coloring in hereditary classes of
  graphs.
\newblock In Michael~A. Bender, Ola Svensson, and Grzegorz Herman, editors,
  {\em 27th Annual European Symposium on Algorithms, {ESA} 2019, September
  9-11, 2019, Munich/Garching, Germany.}, volume 144 of {\em LIPIcs}, pages
  31:1--31:15. Schloss Dagstuhl - Leibniz-Zentrum f{\"{u}}r Informatik, 2019.

\bibitem{DBLP:journals/corr/abs-1802-02282}
Maria Chudnovsky, Sophie Spirkl, and Mingxian Zhong.
\newblock Four-coloring {$P_6$}-free graphs. {I.} extending an excellent
  precoloring.
\newblock {\em CoRR}, abs/1802.02282, 2018.

\bibitem{DBLP:journals/corr/abs-1802-02283}
Maria Chudnovsky, Sophie Spirkl, and Mingxian Zhong.
\newblock Four-coloring {$P_6$}-free graphs. {II.} finding an excellent
  precoloring.
\newblock {\em CoRR}, abs/1802.02283, 2018.

\bibitem{DBLP:conf/soda/SpirklCZ19}
Maria Chudnovsky, Sophie Spirkl, and Mingxian Zhong.
\newblock Four-coloring {$P_6$}-free graphs.
\newblock In Chan \cite{DBLP:conf/soda/2019}, pages 1239--1256.

\bibitem{doi:10.1142/S0129054118500168}
Andreas Darmann, Janosch Döcker, and Britta Dorn.
\newblock The monotone satisfiability problem with bounded variable
  appearances.
\newblock {\em International Journal of Foundations of Computer Science},
  29(06):979--993, 2018.

\bibitem{Ed86}
Keith Edwards.
\newblock The complexity of colouring problems on dense graphs.
\newblock {\em Theor. Comput. Sci.}, 43:337--343, 1986.

\bibitem{EST14}
Jessica Enright, Lorna Stewart, and G{\'{a}}bor Tardos.
\newblock On list coloring and list homomorphism of permutation and interval
  graphs.
\newblock {\em {SIAM} J. Discrete Math.}, 28(4):1675--1685, 2014.

\bibitem{FEDER1998236}
Tomas Feder and Pavol Hell.
\newblock List homomorphisms to reflexive graphs.
\newblock {\em J. Comb. Theory, Ser. {B}}, 72(2):236 -- 250, 1998.

\bibitem{DBLP:journals/combinatorica/FederHH99}
Tom{\'{a}}s Feder, Pavol Hell, and Jing Huang.
\newblock List homomorphisms and circular arc graphs.
\newblock {\em Combinatorica}, 19(4):487--505, 1999.

\bibitem{DBLP:journals/jgt/FederHH03}
Tom{\'{a}}s Feder, Pavol Hell, and Jing Huang.
\newblock Bi-arc graphs and the complexity of list homomorphisms.
\newblock {\em Journal of Graph Theory}, 42(1):61--80, 2003.

\bibitem{DBLP:journals/dam/FederHH09}
Tom{\'{a}}s Feder, Pavol Hell, and Jing Huang.
\newblock Extension problems with degree bounds.
\newblock {\em Discrete Applied Mathematics}, 157(7):1592--1599, 2009.

\bibitem{FHKN05}
Tom{\'{a}}s Feder, Pavol Hell, Sulamita Klein, Loana~Tito Nogueira, and
  F{\'{a}}bio Protti.
\newblock List matrix partitions of chordal graphs.
\newblock {\em Theor. Comput. Sci.}, 349(1):52--66, 2005.

\bibitem{DBLP:conf/stoc/FederV93}
Tom{\'{a}}s Feder and Moshe~Y. Vardi.
\newblock Monotone monadic {SNP} and constraint satisfaction.
\newblock In S.~Rao Kosaraju, David~S. Johnson, and Alok Aggarwal, editors,
  {\em Proceedings of the Twenty-Fifth Annual {ACM} Symposium on Theory of
  Computing, May 16-18, 1993, San Diego, CA, {USA}}, pages 612--622. {ACM},
  1993.

\bibitem{DBLP:journals/dm/GalluccioHN00}
Anna Galluccio, Pavol Hell, and Jaroslav Nesetril.
\newblock The complexity of \emph{H}-colouring of bounded degree graphs.
\newblock {\em Discrete Mathematics}, 222(1-3):101--109, 2000.

\bibitem{DBLP:journals/jgt/GolovachJPS17}
Petr~A. Golovach, Matthew Johnson, Dani{\"{e}}l Paulusma, and Jian Song.
\newblock A survey on the computational complexity of coloring graphs with
  forbidden subgraphs.
\newblock {\em Journal of Graph Theory}, 84(4):331--363, 2017.

\bibitem{DBLP:journals/iandc/GolovachPS14}
Petr~A. Golovach, Dani{\"{e}}l Paulusma, and Jian Song.
\newblock Closing complexity gaps for coloring problems on {$H$}-free graphs.
\newblock {\em Inf. Comput.}, 237:204--214, 2014.

\bibitem{GORSSS18}
Carla Groenland, Karolina Okrasa, Paweł Rzążewski, Alex Scott, Paul Seymour,
  and Sophie Spirkl.
\newblock H-colouring pt-free graphs in subexponential time.
\newblock {\em Discrete Applied Mathematics}, 267:184 -- 189, 2019.

\bibitem{DBLP:journals/corr/GrzesikKPP17}
Andrzej Grzesik, Tereza Klimo\v{s}ov\'{a}, Marcin Pilipczuk, and Michal
  Pilipczuk.
\newblock Polynomial-time algorithm for maximum weight independent set on
  {$P_6$}-free graphs.
\newblock {\em CoRR}, abs/1707.05491, 2017.

\bibitem{DBLP:conf/soda/GrzesikKPP19}
Andrzej Grzesik, Tereza Klimo\v{s}ova, Marcin Pilipczuk, and Michal Pilipczuk.
\newblock Polynomial-time algorithm for maximum weight independent set on
  {$P_6$}-free graphs.
\newblock In Chan \cite{DBLP:conf/soda/2019}, pages 1257--1271.

\bibitem{HN04}
Pavol Hell and Jaroslav Nesetril.
\newblock {\em Graphs and Homomorphisms}.
\newblock Oxford University Press, jul 2004.

\bibitem{HN90}
Pavol Hell and Jaroslav Ne\v{s}et\v{r}il.
\newblock On the complexity of {$H$}-coloring.
\newblock {\em J. Comb. Theory, Ser. {B}}, 48(1):92--110, 1990.

\bibitem{HKLSS10}
Ch{\'{\i}}nh~T. Ho{\`{a}}ng, Marcin Kami\'nski, Vadim~V. Lozin, Joe Sawada, and
  Xiao Shu.
\newblock Deciding $k$-colorability of {$P_5$}-free graphs in polynomial time.
\newblock {\em Algorithmica}, 57(1):74--81, 2010.

\bibitem{Ho81}
Ian Holyer.
\newblock The {NP}-completeness of edge-coloring.
\newblock {\em {SIAM} J. Comput.}, 10(4):718--720, 1981.

\bibitem{Hu16}
Shenwei Huang.
\newblock Improved complexity results on $k$-coloring {$P_t$}-free graphs.
\newblock {\em Eur. J. Comb.}, 51:336--346, 2016.

\bibitem{ImpagliazzoPaturi}
Russell Impagliazzo and Ramamohan Paturi.
\newblock On the complexity of {$k$-SAT}.
\newblock {\em Journal of Computer and System Sciences}, 62(2):367 -- 375,
  2001.

\bibitem{DBLP:journals/jcss/ImpagliazzoPZ01}
Russell Impagliazzo, Ramamohan Paturi, and Francis Zane.
\newblock Which problems have strongly exponential complexity?
\newblock {\em J. Comput. Syst. Sci.}, 63(4):512--530, 2001.

\bibitem{Kr93}
Jan Kratochv\'il.
\newblock Precoloring extension with fixed color bound.
\newblock {\em Acta Mathematica Universitatis Comenianae. New Series}, 62, 01
  1993.

\bibitem{DBLP:journals/dm/LaroseL13}
Benoit Larose and Adrien Lema{\^{\i}}tre.
\newblock List-homomorphism problems on graphs and arc consistency.
\newblock {\em Discrete Mathematics}, 313(22):2525--2537, 2013.

\bibitem{KL07}
Vadim~V. Lozin and Marcin Kami\'nski.
\newblock Coloring edges and vertices of graphs without short or long cycles.
\newblock {\em Contributions to Discrete Mathematics}, 2(1), 2007.

\bibitem{DBLP:conf/stacs/OkrasaR21}
Karolina Okrasa and Pawel Rzazewski.
\newblock Complexity of the list homomorphism problem in hereditary graph
  classes.
\newblock In Markus Bl{\"{a}}ser and Benjamin Monmege, editors, {\em 38th
  International Symposium on Theoretical Aspects of Computer Science, {STACS}
  2021, March 16-19, 2021, Saarbr{\"{u}}cken, Germany (Virtual Conference)},
  volume 187 of {\em LIPIcs}, pages 54:1--54:17. Schloss Dagstuhl -
  Leibniz-Zentrum f{\"{u}}r Informatik, 2021.

\bibitem{DBLP:conf/focs/2017}
Chris Umans, editor.
\newblock {\em 58th {IEEE} Annual Symposium on Foundations of Computer Science,
  {FOCS} 2017, Berkeley, CA, USA, October 15-17, 2017}. {IEEE} Computer
  Society, 2017.

\bibitem{WS01}
Gerhard~J. Woeginger and Jir{\'{\i}} Sgall.
\newblock The complexity of coloring graphs without long induced paths.
\newblock {\em Acta Cybern.}, 15(1):107--117, 2001.

\bibitem{DBLP:conf/focs/Zhuk17}
Dmitriy Zhuk.
\newblock A proof of {CSP} dichotomy conjecture.
\newblock In Umans \cite{DBLP:conf/focs/2017}, pages 331--342.

\end{thebibliography}
\bibliographystyle{plain}

\end{document}